\documentclass[12pt]{iopart}
\usepackage{indentfirst}
\usepackage{graphicx,color}
\usepackage{subfigure}
\usepackage{psfrag}
\usepackage[hang, small, bf, margin=20pt, tableposition=top]{caption}
\usepackage[T1]{fontenc}
\usepackage[square, comma, numbers, sort&compress]{natbib}

\newfont{\bbold}{msbm10 scaled\magstep1}
\newcommand{\bbf}[1]{\mbox{{\bbold #1}}}

\newcommand{\spc}{\quad}

\newcommand{\tbar}{\overline}

\newcommand{\lft}{\left}
\newcommand{\rit}{\right}
\newcommand{\clyd}{\partial}
\newcommand{\mcal}{\mathcal}

\newcommand{\eee}{\varepsilon}

\newcommand{\w}{\omega}
\newcommand{\om}{\Omega}
\newcommand{\ld}{\lambda}

\newcommand{\degrees}{\mbox{$^\circ$}}
\newcommand{\sech}{\mathrm{sech}}
\newcommand{\dd}{{\mathrm d}}
\newcommand{\ee}{{\mathrm e}}
\newcommand{\ii}{{\mathrm i}}
\newcommand{\sfr}[2]{\mbox{$\frac{#1}{#2}$}}
\newcommand{\shalf}{\sfr{1}{2}}
\newcommand{\elo}{E_0}
\newcommand{\wbr}{\mcal{W}_{\mathrm{br}}}

\newenvironment{eqn}{\begin{equation}}{\end{equation}}

\newcommand{\itc}{\emph}
\newcommand{\str}{\textrm}
\newcommand{\mb}{\mathbf}

\newcommand{\fns}{\footnotesize}
\newcommand{\ssz}{\scriptsize}

\newcommand{\lbl}{\label}
\makeatletter
\@addtoreset{equation}{section}
\makeatother
\renewcommand{\theequation}{\arabic{section}.\arabic{equation}}

\begin{document}

\submitto{\JPA}

\title{Discrete breathers in a two-dimensional 
hexagonal Fermi-Pasta-Ulam lattice}
\author{Imran A Butt and Jonathan A D Wattis}
\address{Theoretical Mechanics,  School of Mathematical Sciences,
University of Nottingham, University Park, Nottingham, NG7 2RD,
UK} \eads{\mailto{imran.butt@maths.nott.ac.uk},
\mailto{jonathan.wattis@nottingham.ac.uk}}

\begin{abstract} \lbl{abs}
We consider a two-dimensional Fermi-Pasta-Ulam (FPU) lattice 
with hexagonal symmetry.  Using asymptotic methods based on 
small amplitude ansatz, at third order we obtain a reduction to a 
cubic nonlinear Schr{\"o}dinger equation (NLS) for the breather 
envelope.  However, this does not support stable soliton solutions, 
so we pursue a higher-order analysis yielding a generalised NLS, 
which includes known stabilising terms.   We present numerical 
results which suggest that long-lived stationary and moving breathers 
are supported by the lattice. We find breather solutions which move 
in an arbitrary direction, an ellipticity criterion for the wavenumbers 
of the carrier wave, asymptotic estimates for the breather energy, 
and a minimum threshold energy below which breathers cannot be 
found.  This energy threshold is maximised for stationary breathers, 
and becomes vanishingly small near the boundary of the elliptic domain 
where breathers attain a maximum speed.  Several of the results obtained 
are similar to those obtained for the square FPU lattice (Butt \& Wattis, 
{\em J Phys A}, {\bf 39}, 4955, (2006)),  though we find that the square 
and hexagonal lattices exhibit different properties in regard to the 
generation of harmonics, and the isotropy of the generalised NLS equation.  
\end{abstract}

\pacs{05.45.-a, 05.45.Yv \\[2ex]This version: \today}

\section{Introduction} \lbl{hexint}

Discrete breathers are time-periodic and spatially localised exact
solutions of translationally invariant nonlinear lattices.  For a
brief review of some general properties of breathers in
higher-dimensional systems, see our earlier work \citep{buts06}.  
In particular, it is known that while some fundamental properties
such as the existence of breathers are not affected by lattice
dimension (see Flach \etal \citep{flat94} and Mackay and Aubry
\citep{mac94}), other properties are affected profoundly, for
instance, the energy properties of breathers (see Flach \etal
\citep{fla97}, Kastner \citep{kas04} and Weinstein \citep{wei99}).
In this paper we investigate how the symmetry of the lattice 
influences the properties of discrete breathers found therein.

In Hamiltonian systems, stationary breathers occur in one-parameter 
families.  For a certain class of Hamiltonian systems, a critical spatial 
dimension $d_c$ exists such that for systems with $d \geq d_c$, 
there exists a positive lower bound on the energy of a breather 
family, and the breather energies do {\em not} approach zero even 
as the amplitude tends to zero.   For lattices with dimension 
$d < d_c$, there is no positive lower bound on the energy of 
breathers.  In other words, the energy of a family of breathers goes 
to zero with amplitude, and breathers of arbitrarily small energy can 
be found.    The critical dimension $d_c$ is typically two.  A small 
amplitude expansion yields the NLS reduction to $\ii F_T+\Delta F 
+ \kappa |F|^{2\sigma}F=0$ which has a critical dimension of  
$d_c=2/\sigma$, with blow-up in NLS occurring when $\kappa>0$ 
and $d>d_c$.   For typical lattice potentials, $\sigma=1$ again 
confirming the critical dimension of $d_c=2$.

Marin, Eilbeck and Russell have performed extensive numerical 
investigations of breather dynamics in two-dimensional lattices 
\citep{mar98,mar00,mar01}.  Their results suggest that moving 
breather modes exist (or are at least extremely long-lived), and 
that the lattice exhibits a strong directional preference whereby 
breathers can only move along symmetries of the lattice, and in 
no other direction.  Such quasi-one-dimensional behaviour is also 
observed in higher-dimensional lattices.

The work in this paper follows on from our earlier study 
\citep{buts06} of breathers in a two-dimensional lattice with square 
rotational symmetry;  hereupon referred to as the ``square'' lattice.  
This lattice has $C_4$ symmetry, by which we mean that a rotation 
through any multiple of $2\pi/4 = \pi/2$ about an axis perpendicular to 
the lattice plane through a lattice site maps the lattice onto itself.  
In \cite{buts06}, using the semi-discrete multiple-scale method 
(see Remoissenet \citep{rem85}), we determined an approximate 
form for (as well as the properties of) small 
amplitude breathers in a two-dimensional square Fermi-Pasta-Ulam 
(FPU) lattice \citep{fer40}.  We found a third-order analysis to be 
inadequate, since the partial differential equation obtained at this order 
describing the breather envelope exhibits \itc{blow-up}.  To
overcome this, we incorporated higher-order effects in the model,
thereby obtaining a modified partial differential equation which
includes known stabilising terms. From this, we determined regions
of parameter space where breather solutions are expected.
Numerical simulations supported the results of our analysis, and
suggested that, in contrast to the two-component lattices studied
by Marin \etal\ \citep{mar98,mar00,mar01} (that is, with two
degrees of freedom at each lattice site), there is no restriction upon 
the permitted directions of travel within the one-component square 
FPU lattice. We also found asymptotic estimates for the breather 
energy which confirmed the existence of a minimum threshold 
energy, in agreement with the work of Flach \etal\ \citep{fla97}.

\begin{figure}[htbp]
\begin{center}
\resizebox{5in}{!}{\includegraphics{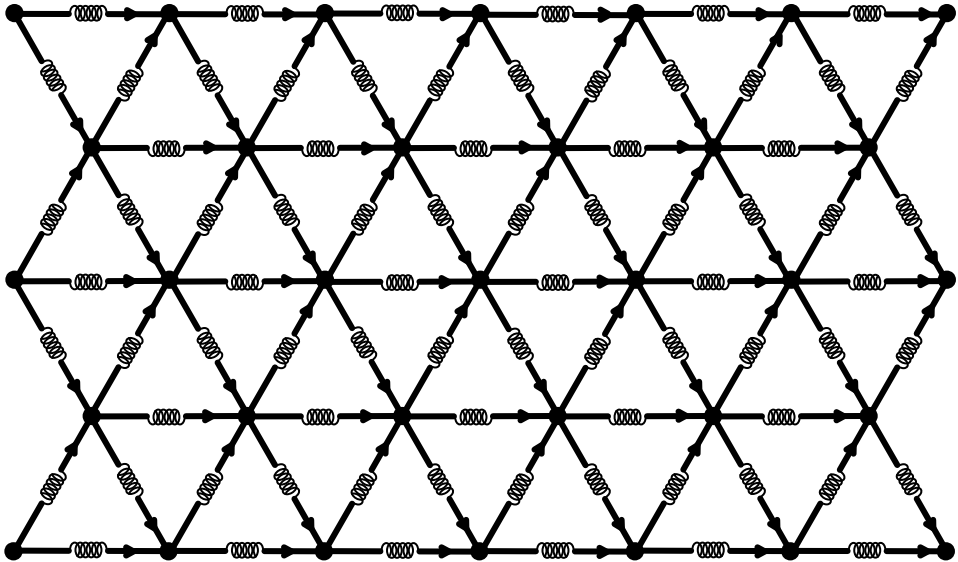}}
\caption{The 2D hexagonal electrical transmission lattice (HETL).}
\lbl{hexetl}
\end{center}
\end{figure}

In this paper, we consider a hexagonal electrical transmission lattice 
(HETL).  This two-dimensional network possesses $C_6$ (or hexagonal) 
rotational symmetry.  That is, rotation through any multiple of the angle 
$2\pi/6 = \pi/3$ about a lattice site maps the lattice onto itself.  The HETL 
is shown in \Fref{hexetl}, pictured from a point vertically above the plane 
of the lattice.  We note from \Fref{hexetl} that geometrically, the HETL is 
an arrangement of tessellating triangles (not hexagons).   Nevertheless, the 
arrangement in \Fref{hexetl} is referred to as ``hexagonal'' rather than 
``triangular'' since these descriptions refer to its symmetry properties and 
not to the geometrical shapes which comprise the array (not all authors 
follow this convention).  One might expect the analysis for the hexagonal 
lattice to be more involved, since it is geometrically more complicated in 
having more links emanating from each node.  However, the hexagonal 
symmetry results in greater isotropy and hence simpler equations than 
those obtained for the square lattice.

We derive the equations of motion and demonstrate a Hamiltonian
formalism in \Sref{hexder}.  In Sections \ref{hexsym} and
\ref{hexasymm}, we present two cases for which the hexagonal FPU
lattice equations can be reduced to a two-dimensional nonlinear
Schr{\"o}dinger (NLS) equation with cubic nonlinearity.  We
consider lattices with a symmetric interaction potential, in which
case reduction to a cubic NLS equation can be performed for moving
breathers.  We find an ellipticity criterion for the wavenumbers
of the carrier wave in \Sref{hexdell}.  A reduction to the cubic NLS
equation can also be carried out for lattices with an asymmetric
potential, provided we consider only stationary breathers.

As expected we find a minimum energy below which breathers cannot 
exist in the hexagonal FPU lattice.  We find that the
energy threshold is dependent upon the wavevector of the carrier
wave. It is maximised for stationary breathers, and becomes
arbitrarily small near the boundary of the elliptic domain.

The cubic NLS equation admits only unstable Townes solitons.  Hence, 
in \Sref{hex5}, we extend our asymptotic analysis to higher order and 
find an isotropic generalised NLS equation which incorporates known 
stabilising terms.  Our analytic work is supplemented by numerical simulations 
presented in \Sref{hexnumerics}, which suggest that long-lived stationary
and moving breather modes are supported by the system.
In \Sref{hdisc}, we discuss the results obtained in this paper.

\section{A two-dimensional hexagonal Fermi-Pasta-Ulam lattice}
\lbl{hexfpulat}

\subsection{Preliminaries} \lbl{hexpre}

Before we derive the equations governing the HETL, we describe our 
scheme for indexing the lattice nodes and our choice of basis vectors.  We 
introduce a rectangular lattice with basis vectors $\mcal{B}~=~\{ \mb{i'} \, , 
\, \mb{j'} \}$, where $\mb{i'} = \mb{i} = [1,0]^T$ and $\mb{j'} = [0,h]^T$ 
($\mb{j}$ being $[0,1]^T$), illustrated in \Fref{sqbase}.  We use only half of 
the $(m,n)$ indices, namely, those for which the sum $m+n$ is even.  We 
choose an origin with coordinates $(0,0)$;  the position of the site $(m,n)$ 
is $m\mb{i'} + n\mb{j'}$.  In order for the hexagonal lattice to be regular, 
we specify $h=\sqrt{3}$.

\begin{figure}[htbp]
\begin{center}
\vspace{-1.0cm}
\psfrag{0}{{\scriptsize $(m,n)$}}
\psfrag{1}{{\scriptsize $(m+2,n)$}}
\psfrag{2}[B][]{{\scriptsize $(m+1,n+1)$}}
\psfrag{3}[B][]{{\scriptsize $(m-1,n+1)$}}
\psfrag{4}[B][r]{{\scriptsize $(m-2,n)$}}
\psfrag{5}[B][b]{{\scriptsize $(m-1,n-1)$}}
\psfrag{6}[B][b]{{\scriptsize $(m+1,n-1)$}}
\psfrag{i}{{\footnotesize $\mb{i'}$}}
\psfrag{j}{{\footnotesize $\mb{j'}$}}
\resizebox{4in}{!}{\includegraphics{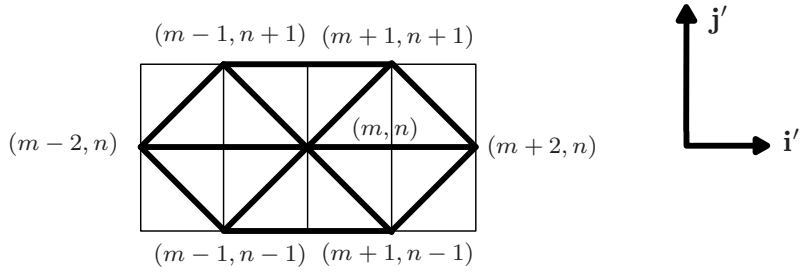}}
\caption{Labelling of nodes in the HETL with basis 
$ \mcal{B} = \{ \mb{i'} \, , \, \mb{j'} \}$.}
\lbl{sqbase}
\end{center}
\end{figure}

At every node of the HETL lies a nonlinear capacitor (not shown in 
\Fref{hexetl}), and between every node and each of its six nearest 
neighbours is a linear inductor.  An enlarged view of the area surrounding 
the capacitor at $(m,n)$ is shown in \Fref{hexunitdir}, where the capacitor 
is visible.  The variable $V_{m,n}$ denotes the voltage across the 
capacitor $(m,n)$ and $Q_{m,n}$ denotes the total charge stored on the 
capacitor at $(m,n)$.  Also, $I_{m,n}$, $J_{m,n}$ and $K_{m,n}$ are the 
currents through the inductors immediately to the right of site $(m,n)$ in 
the directions $\mb{e_i} = [2,0]^T$, $\mb{e_j} = [1,-\sqrt{3}]^T$ and 
$\mb{e_k}=[1,\sqrt{3}]^T$ respectively, as illustrated in \Fref{hexunitdir}.

\begin{figure}[htbp]
\begin{center}
\vspace{1.3cm}
\psfrag{a}{{\large $V_{m,n}$}}
\psfrag{b}[B][]{{\large $V_{m+1,n-1}$}}
\psfrag{c}[B][]{{\large $V_{m+2,n}$}}
\psfrag{d}{{\large $V_{m+1,n+1}$}}
\psfrag{e}{{\large $I_{m,n}$}}
\psfrag{f}{{\large $J_{m,n}$}}
\psfrag{g}{{\large $K_{m,n}$}}
\psfrag{i}{{\normalsize $\mb{e_i}$}}
\psfrag{j}{{\normalsize $\mb{e_j}$}}
\psfrag{k}{{\normalsize $\mb{e_k}$}}
\resizebox{5.5in}{!}{\includegraphics{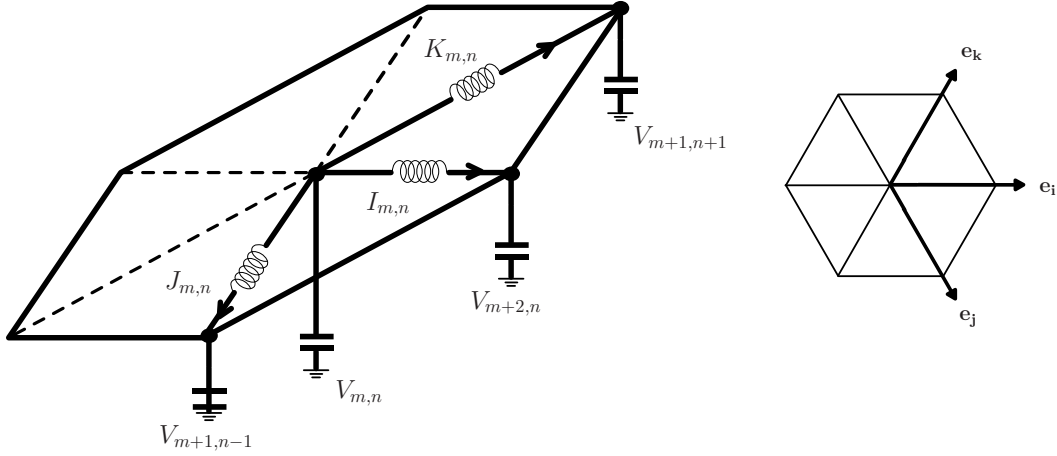}}
\caption{Enlarged view of the HETL at site $(m,n)$.}
\lbl{hexunitdir}
\end{center}
\end{figure}

\subsection{Derivation of model equations} \lbl{hexder}

To derive the equations relating current, charge and voltage in the lattice
we apply Kirchoff's law 
\begin{eqnarray}
V_{m+2,n} - V_{m,n} & = -L \frac{\dd I_{m,n}}{\dd t}, \lbl{hki} \\[1ex]
V_{m+1,n-1} - V_{m,n} & = -L \frac{\dd J_{m,n}}{\dd t}, \lbl{hkj} \\[1ex]
V_{m+1,n+1} - V_{m,n} & = -L \frac{\dd K_{m,n}}{\dd t}, \lbl{hkk}
\end{eqnarray}
where the inductance $L$ is constant.  Conservation of charge gives
\begin{eqn} \lbl{hcc}
I_{m-2,n}-I_{m,n} + J_{m-1,n+1}-J_{m,n} + K_{m-1,n-1}-K_{m,n} =
\frac{\dd Q_{m,n}}{\dd t}.
\end{eqn}
Differentiating \eref{hcc} with respect to time, and then using 
\eref{hki}--\eref{hkk} to find $\dot{I}_{m-2,n}$, $\dot{J}_{m-1,n+1}$ 
and $\dot{K}_{m-1,n-1}$, we have  
\begin{eqnarray}
L \frac{\dd^2 Q_{m,n}}{\dd t^2} & = & \spc (V_{m+2,n} - 2 V_{m,n} + 
V_{m-2,n}) + (V_{m+1,n-1} - 2 V_{m,n} + V_{m-1,n+1}) \nonumber \\ & & 
+ (V_{m+1,n+1} - 2 V_{m,n} + V_{m-1,n-1}). 
\lbl{heqvqfull} \end{eqnarray}
Equation \eref{heqvqfull} may be written in the abbreviated form
$L \ddot{Q}_{m,n} = (\delta^2_I + \delta^2_J + \delta^2_K)V_{m,n}$,
where the centred second-difference operators are defined by
\begin{eqnarray}
\lbl{hdi} \delta^2_I A_{m,n} & = A_{m+2,n}   - 2 A_{m,n} + A_{m-2,n},\\[1ex]
\lbl{hdj} \delta^2_J A_{m,n} & = A_{m+1,n-1} - 2 A_{m,n} + A_{m-1,n+1},\\[1ex]
\lbl{hdk} \delta^2_K A_{m,n} & = A_{m+1,n+1} - 2 A_{m,n} + A_{m-1,n-1} . 
\end{eqnarray}
Here, $A_{m,n}$ is an arbitrary quantity referenced by two indices; $\delta^2_I$, 
$\delta^2_J$ and $\delta^2_K$ are centred second-difference operators in 
the directions of $\mb{e_i}$, $\mb{e_j}$ and $\mb{e_k}$ respectively.   
Since the voltage $V_{m,n}$ is known in terms of the charge, we reformulate 
\eref{heqvqfull} in terms of the single quantity $Q_{m,n}$.  We invert the 
capacitor's nonlinear charge-voltage relationship $Q=VC(V)$ (see 
equations (2.9)--(2.14) of \cite{buts06} for details), to obtain 
\begin{equation} 
V(Q) = ( Q + a Q^2 + b Q^3 + c Q^4 + d Q^5 ) / C_0 , 
\lbl{hqvexp} \end{equation}
where $C_0=C(0)$.  Hence, the HETL equations \eref{heqvqfull} can 
be written as 
\begin{eqn} \lbl{heqq}
\frac{\dd^2 Q_{m,n}}{\dd t^2} = (\delta^2_I + \delta^2_J + \delta^2_K) 
\lft[Q_{m,n}+aQ_{m,n}^2+bQ_{m,n}^3+cQ_{m,n}^4+dQ_{m,n}^5\rit], 
\end{eqn}
where $m,n \in {\bbf{Z}}$ and, by rescaling the time variable, we set 
$LC_0=1$ without loss of generality.   Thus we have shown that the 
equation governing charge in the HETL \eref{heqq} is a two-dimensional 
analogue of the Fermi-Pasta-Ulam equation 
\begin{equation}
\frac{\dd^2 Q_j}{\dd t^2} = W'(Q_{j+1}) - 2 W'(Q_j) + W'(Q_{j-1}) , 
\end{equation}
which is a Hamiltonian system that can be derived from both 
\begin{equation}
H = \sum_j \shalf \pi_j^2 + W(\phi_{j+1}-\phi_j) , \;\;\; {\rm and} \;\;\; 
\widetilde H = \sum_j \shalf (P_{j+1}-P_j)^2 + W(Q_j) , 
\end{equation}
where $Q_j=\phi_{j+1}-\phi_j$.

The lattice equations \eref{heqq} can be derived from the Hamiltonian 
\begin{eqnarray}
\widetilde{H} & =\sum_{m,n} & 
\mbox{$\frac{1}{2}$} (P_{m+2,n}-P_{m,n})^2 + 
\mbox{$\frac{1}{2}$} (P_{m+1,n-1} - P_{m,n})^2 \nonumber \\ & & + 
\mbox{$\frac{1}{2}$} (P_{m+1,n+1} - P_{m,n})^2 + \Upsilon (Q_{m,n}), 
\lbl{hham} \end{eqnarray}
where $\Upsilon (Q_{m,n})$ satisfies $\Upsilon\!\ '(Q_{m,n}) =
V(Q_{m,n})$ given in \eref{hqvexp}, hence   
\begin{eqn}
\Upsilon(Q) = \mbox{$\frac{1}{2}$} Q^2 + 
\mbox{$\frac{1}{3}$} a Q^3 + \mbox{$\frac{1}{4}$} b Q^4 + 
\mbox{$\frac{1}{5}$} c Q^5 + \mbox{$\frac{1}{6}$} d Q^6 .
\end{eqn} 
We describe potentials which satisfy $\Upsilon(-Q)=\Upsilon(Q)$ (that is, 
$a=c=0$) as `symmetric'.  The variables $P_{m,n}$ and $Q_{m,n}$ are 
conjugate momenta and displacement variables of the system; and 
eliminating $P_{m,n}$ from the equations 
\begin{eqn} \lbl{hhameq}
\frac{\dd Q_{m,n}}{\dd t} = -(\delta^2_I + \delta^2_J  +
\delta^2_K)P_{m,n}, \qquad
\frac{\dd P_{m,n}}{\dd t} = -\Upsilon\!\ '(Q_{m,n}) , 
\end{eqn}
yields (\ref{heqq}). In \Sref{heeh}, we derive expressions for 
the energy of breathers given the small amplitude solutions 
which are obtained in the next section.

\subsection{Asymptotic analysis} \lbl{haa}

We apply the method of multiple-scales to determine an approximate 
analytic form for small amplitude breather solutions of \eref{heqq}, with 
slowly varying envelope.  We introduce new variables defined by
 \begin{eqn} \lbl{hexvars}
 X=\eee m,\spc Y=\eee h n,\spc\tau=\eee t\spc\str{and}\spc T=\eee^2 t;
 \end{eqn}
note the presence of the scaling factor $h$ in the definition of $Y$.  
We seek solutions of \eref{heqq} of the form
\begin{eqnarray} \fl 
Q_{m,n}(t) = \hspace*{-3mm} & \!\!\!\! \eee \ee^{\ii \psi} F(X,Y,\tau,T) + 
\eee^2 G_0(X,Y,\tau,T) +
\eee^2 \ee^{\ii \psi} G_1(X,Y,\tau,T) + \nonumber \\ & \!\!\!\! 
\eee^2 \ee^{2\ii \psi} G_2(X,Y,\tau,T) + \eee^3 H_0 (X,Y,\tau,T) +
\eee^3 \ee^{\ii \psi} H_1(X,Y,\tau,T) + \nonumber \\
& \!\!\!\!  \eee^3 \ee^{2\ii \psi} H_2(X,Y,\tau,T)
+ \eee^3\ee^{3\ii \psi}H_3(X,Y,\tau,T) + \eee^4\ee^{\ii\psi}I_1(X,Y,\tau,T)
+ \nonumber\\ & \!\!\!\!  
\eee^4 \ee^{2\ii \psi} I_2(X,Y,\tau,T) + \eee^4 \ee^{3\ii\psi} I_3(X,Y,\tau,T)
+ \eee^4\ee^{4\ii \psi} I_4(X,Y,\tau,T) + \nonumber \\
& \!\!\!\! \eee^5\ee^{\ii \psi} J_1(X,Y,\tau,T)+ \cdots + \mathrm{c.c.},
\lbl{hexanz} \end{eqnarray}
where the phase $\psi$ of the carrier wave is given by $km+lhn+\w t$ (once 
again noting the extra factor $h$), and $\mb{k}=[k,l]^T$ and $\w(\mb{k})$ 
are its wavevector and temporal frequency respectively.  We substitute the 
ansatz \eref{hexanz} into the lattice equations \eref{heqq} and equate 
coefficients of each harmonic frequency at each order of $\eee$.  After 
much simplification, this yields the following system of equations: \\
\\
$\mathcal{O}(\eee \ee^{\ii\psi})$:
\begin{eqn} \lbl{hexdisp}
\w^2 F = 4\sin^2(k) F + 4\sin^2 \lft( \frac{k+lh}{2} \rit) F + 
4\sin^2 \lft( \frac{k-lh}{2} \rit) F,
\end{eqn}
$\mathcal{O}(\eee^2 \ee^{\ii\psi})$:
\begin{eqn} \lbl{hexvel}
\w F_{\tau} = 2\sin k [2\cos k + \cos(lh)]F_X + 2h\cos k \sin(lh) F_Y,
\end{eqn}
$\mcal{O}(\eee^2 \ee^{2\ii\psi})$:
\begin{eqn} \lbl{hexg2}
\w^2 G_2 = [\sin^2(2k) + \sin^2(k+lh) + \sin^2(k-lh)](G_2 + a F^2),
\end{eqn}
$\mcal{O}(\eee^3 \ee^{\ii\psi})$:
\begin{eqnarray}
\fl 2\ii \w F_T + F_{\tau\tau}  & 
= & [4\cos(2k) + 2\cos k \cos(lh)] F_{XX} + 2h^2\cos k \cos(lh)
F_{YY} \nonumber \\
& & - 4h\sin k \sin(lh)F_{XY} \nonumber \\
& & - \ 8a \lft[ \sin^2(k) + \sin^2 \lft( \frac{k+lh}{2} \rit)
+ \sin^2 \lft( \frac{k-lh}{2} \rit) \rit] [F(G_0+\tbar{G}_0)+\tbar{F}G_2] 
\nonumber\\
& & - \ 12 b \lft[ \sin^2(k) + \sin^2 \lft( \frac{k+lh}{2} \rit)
+ \sin^2 \lft( \frac{k-lh}{2} \rit) \rit] |F|^2 F,  \lbl{hexnls}
\end{eqnarray}
$\mcal{O}(\eee^3 \ee^{3\ii\psi})$:
\begin{eqn} \fl 
9\w^2H_3 =  4 \lft[ \sin^2(3k) + \sin^2 \lft( \frac{3k+3lh}{2} \rit) + 
\sin^2 \lft( \frac{3k-3lh}{2} \rit) \rit] \left( H_3 + 2aFG_2+bF^3 \right), 
\lbl{hexh3} \end{eqn}
$\mcal{O}(\eee^4 \ee^0)$:
\begin{eqn} \lbl{hexfo}
G_{0\tau\tau} = 6\, G_{0XX} + 2h^2 G_{0YY} + 
a \! \lft[ 6 \lft( |F|^2 \rit)_{XX} + 2h^2 \lft(|F|^2 \rit)_{YY} \rit] .
\end{eqn}

Though each equation plays a similar role to its counterpart in the square 
lattice, equations \eref{hexdisp}--\eref{hexfo} are more complicated. 
Equation \eref{hexdisp} is the dispersion relation for the system \eref{heqq}.  
Since we are interested only in solutions for which $F \neq 0$, \eref{hexdisp} yields 
\begin{eqn} \lbl{hexdisp2} 
\w^2  = 4\sin^2(k)  + 4\sin^2 \lft( \frac{k+lh}{2} \rit) + 
4\sin^2 \lft( \frac{k-lh}{2} \rit),
\end{eqn}
which does not simplify significantly.  From equation \eref{hexvel}, 
we determine the velocity of the travelling wave $F$, finding 
\begin{eqn}  \lbl{hexvels}
F(X,Y,\tau,T) \equiv F(Z,W,T),
\end{eqn}
where $Z=X-u\tau$ and $W=Y-v\tau$, and the horizontal and vertical
velocity components ($u$ and $v$) are found to be
\begin{eqn}  \lbl{velsuv}
u = \frac{-2\sin (k)[2\cos(k)+\cos(lh)]}{\w} \spc\str{and}\spc
v = \frac{-2h\cos(k)\sin(lh)}{\w}.
\end{eqn}
Equation \eref{velsuv}, along with \eref{hexdisp2} enables the elimination 
of terms involving $G_1$ from \eref{hexnls} which are not shown.  We 
denote the angle at which the envelope $F$ propagates through the 
lattice by $\Psi$, which is measured from the direction of the basis vector 
$\mb{e_i}$ to the line of travel and hence is given by $\tan^{-1}(v/u)$, 
which in turn depends upon the wavevector  $\mb{k}$.   For both the cases 
that we consider (namely, symmetric and asymmetric interaction potentials), 
we find constraints upon the wavenumbers $k$ and $l$ which affect the 
velocity components $u$ and $v$.   By taking $k=\pi/3$ and  $l=\pi/h$ we 
find $u=v=0$, which corresponds to a static breather; and by choosing 
$k,l$ values which circle this point, we find breathers which can propagate 
in any direction ($0\leq\Psi<2\pi$); in other words, our analysis suggests 
that there is no restriction upon the direction of travel through the lattice.

Our aim is to reduce \eref{hexnls} to a nonlinear
Schr{\"o}dinger (NLS) equation for $F$.  Before this
can be done, the quantities $G_0$ and $G_2$ in \eref{hexnls} must
be found in terms of $F$.  As for the square lattice, it is straightforward
to determine $G_2$ from the algebraic equation \eref{hexg2}.
However, the partial differential equation \eref{hexfo} for $G_0$
can be solved for two special cases only; namely, 
symmetric potentials, in which case the reduction of \eref{hexnls}
to an NLS equation can be completed even for moving breathers, and
also asymmetric potentials, provided we confine our attention to
stationary breathers.  These two cases are considered in Sections 
\ref{hexsym} and \ref{hexasymm} respectively, where we also use 
our breather formulae to generate estimates for the breather energy.

\subsection{Asymptotic estimates for breather energy} \lbl{heeh}

The HETL is a lossless network, meaning that the total electrical energy 
is conserved.  This quantity is related to the Hamiltonian 
(\ref{hham}) by $E=\widetilde{H}/C_0$ and so is given by 
\begin{eqn} \lbl{hexentot}
E = \sum_{m,n} e_{m,n} =  \sum_{m,n} \frac{\Upsilon(Q_{m,n})}{C_0} 
+ \mbox{$\frac{1}{2}$}L \lft( I_{m,n}^2 + J_{m,n}^2 + K_{m,n}^2 \rit) . 
\end{eqn}
The electrical energy $e_{m,n}$ associated with a unit of the lattice is 
(see \Fref{hexunitdir})
\begin{eqn} \lbl{hexenun} 
e_{m,n} = \frac{\Upsilon(Q_{m,n})}{C_0} + 
\mbox{$\frac{1}{2}$}L ( I_{m,n}^2 + J_{m,n}^2 + K_{m,n}^2 ) . 
\end{eqn}
To derive a leading-order estimate for the electrical energy, defined by 
\begin{eqn}  \lbl{hentotlo}
E_0 = \sum_{m,n} e_{m,n}^{(0)}=\sum_{m,n}\frac{Q_{m,n}^2}{2C_0}
+ \frac{L}{2} \lft( I_{m,n}^2 + J_{m,n}^2 + K_{m,n}^2 \rit) , 
\end{eqn}
we use leading-order expressions for each of the terms in the terms in 
the summand of \eref{hexentot}.  The first term is $Q_{m,n}^2/(2C_0)$;  
from \eref{hqvexp}, it follows that $V_{m,n} \sim Q_{m,n}/C_0$, and 
so to leading order, (\ref{hexentot}) agrees with the linear approximation 
to the energy in the capacitor being $QV/2$.   To find leading-order 
expressions for the currents  $I_{m,n}$, $J_{m,n}$ and $K_{m,n}$, we 
use equations \eref{hki}--\eref{hkk}, which imply 
\begin{eqnarray}
Q_{m+2,n} - Q_{m,n} & = - \frac{\dd I_{m,n}}{\dd t}, \lbl{hkiq} \\[1ex]
Q_{m+1,n-1} - Q_{m,n} & = - \frac{\dd J_{m,n}}{\dd t}, \lbl{hkjq} \\[1ex]
Q_{m+1,n+1} - Q_{m,n} & = - \frac{\dd K_{m,n}}{\dd t}, \lbl{hkkq}
\end{eqnarray}
where $LC_0=1$.  The currents are determined by substituting the
expression for the breather $Q_{m,n}$ into \eref{hkiq}--\eref{hkkq} 
and then integrating with respect to time.  We obtain leading-order 
estimates for the energy of moving breathers in systems with symmetric 
potentials Section \ref{hemv}, and in Section \ref{hest}, the energies 
of stationary breathers with asymmetric potentials.

\subsection{The dispersion relation for the HETL} \lbl{wplots}

In this section, we analyse the dispersion relation \eref{hexdisp2} 
for the system \eref{heqq}.  A contour plot of $\w$ against $k$ and 
$l$ is shown in \Fref{wcont}.  Since $w$ is periodic in both $k$ and 
$l$, with period $2\pi$ along the $k$-direction, and $2\pi/h$ in the 
$l$-direction, we consider only $k$ and $l$ such that 
$(k,l) \in \mcal{T}^2 = [0,2\pi] \times [0,2\pi/h]$.

The function $\w$ is minimised, and assumes the value zero, at the
centre of the circular patterns in \Fref{wcont}.  The minima are located 
at $(0,0)$, $(2\pi,0)$, $(2\pi,2\pi/h)$, $(0,2\pi/h)$ and $(\pi,\pi/h)$.  
Plus signs (`$+$') mark the points in $(k,l)$-space at which $\w$ is 
maximised and takes the value $\w=3$.  The maxima lie at the centres 
of the equilateral triangles, the wavevectors corresponding to these 
points are denoted $\mb{k_1}, \ldots, \mb{k_6}$, where
\begin{equation} \begin{array}{ lclcl}
\mb{k_1}=[\pi/3,\pi/h]^T,  && 
\mb{k_2}=[2\pi/3,0]^T,  && 
\mb{k_3}=[4\pi/3,0]^T,    \\ 
\mb{k_4}=[5\pi/3,\pi/h]^T, && 
\mb{k_5}=[4\pi/3,2\pi/h]^T, &&
\mb{k_6}=[2\pi/3,2\pi/h]^T.
\end{array} \end{equation} 
The arrangement of the points corresponding to these wavevectors
in $(k,l)$-space reflects the hexagonal symmetry properties of the
lattice \eref{heqq}.  It may be verified using \eref{velsuv} that
the velocity components $u$ and $v$ are both zero for each of the
wavevectors $\{ \mb{k_1},\ldots, \mb{k_6} \}$.

\begin{figure}[htbp]
\begin{center}
\resizebox{4.5in}{!}{\includegraphics{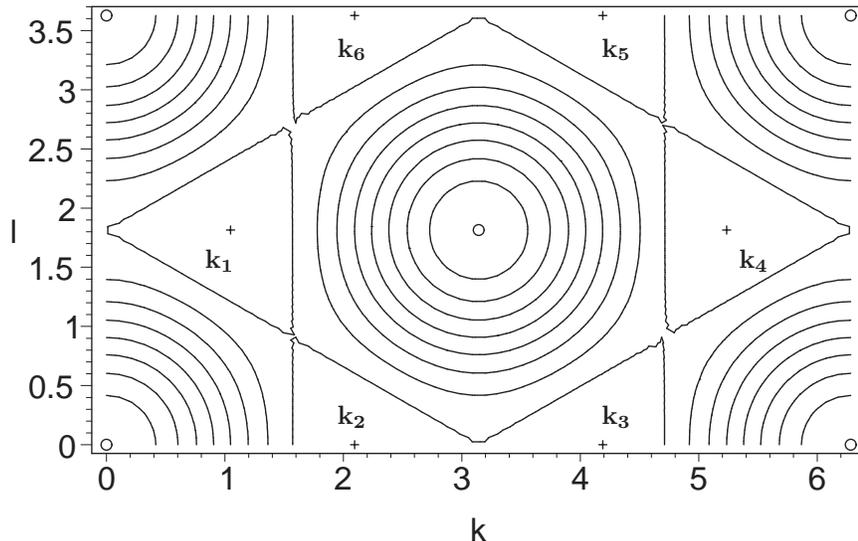}}
\put(-250,103){\footnotesize $\mb{k_1}$}
\put(-200,44){\footnotesize $\mb{k_2}$}
\put(-100,44){\footnotesize $\mb{k_3}$}
\put(-48,103){\footnotesize $\mb{k_4}$}
\put(-100,183){\footnotesize $\mb{k_5}$}
\put(-200,183){\footnotesize $\mb{k_6}$}
\caption{Contour plot of $\w(\mb{k})$; $\w$ attains its maximum 
value of 3 at the points $\mb{k_1},\ldots,\mb{k_6}$, and 
is minimised at the points marked `o', where $\w=0$. }
\lbl{wcont}
\end{center}
\end{figure}

\subsection{Lattices with a symmetric potential} \lbl{hexsym}

In this section, we consider lattices with a symmetric interaction potential, 
where $\Upsilon (Q)$ is even and $\Upsilon\!\ '(Q)$  has odd symmetry. 
This corresponds to $a=c=0$ in \eref{heqq}.  Since there are no even 
harmonics for vibrations controlled by symmetric potentials, it follows 
that $G_0$ and $G_2$ are both zero.  In \eref{hexnls}, the term 
$F_{\tau\tau}$ is eliminated using $F_{\tau\tau} = u^2F_{ZZ} + 2uvF_{ZW} 
+ v^2F_{WW}$ which is derived from \eref{hexvels}.  This 
leads to (\ref{hexnls}) being reduced to an NLS equation 
\begin{eqnarray}
\fl 2\ii\w F_T + \lft[ u^2 - 4\cos(2k) - 2\cos k\cos(lh) \rit] F_{ZZ}
+ \lft[ v^2 - 2h^2\cos k \cos(lh) \rit] F_{WW} \nonumber \\
+ \lft[ 2uv + 4h \sin k \sin(lh) \rit] F_{ZW} + 3b\w^2|F|^2F = 0, 
\lbl{hexrnls} \end{eqnarray}
where the velocities $u$ and $v$ are given by \eref{velsuv}.  By
applying an appropriate change of variables, we remove 
the mixed derivative term from  \eref{hexrnls}, and reduce the
equation to a standard form.  To simplify the appearance of
subsequent expressions, we denote the coefficients of $F_{ZZ}$,
$F_{WW}$ and $F_{ZW}$ by $D_1 = u^2 - 4\cos(2k) - 2\cos
k\cos(lh)$, $D_2 =  v^2 - 2h^2\cos k \cos(lh)$ and $D_3 = 2uv + 4h
\sin k \sin(lh)$ respectively.  A suitable transformation is thus
\begin{eqn} \lbl{hextr}
\xi = \frac{hZ}{\sqrt{D_1}} \spc \str{and} \spc
\eta = \frac{h(2D_1 W - D_3 Z)}{\sqrt{D_1 (4D_1 D_2 - D_3^2)}},
\end{eqn}
which implies \eref{hexrnls} becomes
\begin{eqn} \lbl{hexcan}
2\ii\w F_T + 3\nabla^2 F + 3b\w^2|F|^2F = 0,
\end{eqn}
where the differential operator $\nabla^2 F \equiv F_{\xi\xi} +
F_{\eta\eta}$ is isotropic in the $(\xi,\eta)$ variables.   An approximation 
to the Townes soliton solution of (\ref{hexcan}) is given by equation 
(\ref{vartown}) in the appendix.  Substituting the resulting expression for $F$ 
into \eref{hexanz} yields a leading-order expression for the breather
\begin{eqn} \lbl{hexlo}
Q_{m,n}(t) = 2 \eee \alpha \cos[km+lhn+(\w+\eee^2\lambda)t] 
\ \sech(\beta r) + \mcal{O}(\eee^3),
\end{eqn}
where $\alpha$ and $\beta$ are determined as described by equation 
(\ref{alpbet}) in the appendix, with $D=3/2\w$, $B=3b\w/2$.  Further,
$r=\sqrt{\xi^2+\eta^2}$ is found in terms of the physical discrete
variables $m$ and $n$ by inverting the transformations
\eref{hextr} and reverting back to the variables $Z$ and $W$, using
\begin{eqnarray} 
r^2 &=& \xi^2 + \eta^2 = \frac{ 4 h^2 \eee^2 
( D_2(m\!-\!ut)^2 + D_1(hn\!-\!vt)^2 - D_3(m\!-\!ut)(hn\!-\!vt) )}
{4D_1D_2 -D_3^2}.  \nonumber \\ && \lbl{hexr}
\end{eqnarray}
The terms $D_1$, $D_2$ and $D_3$ are known from \eref{hexrnls}, the 
velocities $u$ and $v$ are given by \eref{velsuv} and $\w$ is given in 
\eref{hexdisp2}.

\subsubsection{Determining the domain of ellipticity. } \lbl{hexdell}

We confine our attention to elliptic NLS equations.  We seek to determine 
the region $\mcal{D}$ of $(k,l)$-parameter space (that is, the two-torus 
$\mcal{T}^2 = [(0,2\pi)]\times [0,2\pi/h]$) where the NLS equation 
\eref{hexrnls} is elliptic. By definition, this equation is elliptic when 
$D_3^2 < 4 D_1 D_2$, where $D_1$, $D_2$ and $D_3$ are given in 
\eref{hexrnls}.  Whilst the region $\mcal{D}$ cannot be specified 
explicitly,  it is simple to find numerically, and is illustrated in  \Fref{econt}.  
Defining the function $e(k,l) = 4D_1(k,l) \! \cdot \! D_2(k,l) - D_3(k,l)^2$, 
we are concerned with the region where $e(k,l)>0$.  The subdomains 
have been labelled $\{ \mcal{D}_1 ,\ldots, \mcal{D}_6\}$ in \Fref{econt}, 
where $\mcal{D} = \mcal{D}_1 \cup \ldots \cup \mcal{D}_6$.

\addtolength{\captionmargin}{-15pt}
\begin{figure}[ht]
\begin{center}
\resizebox{4.5in}{!}{\includegraphics{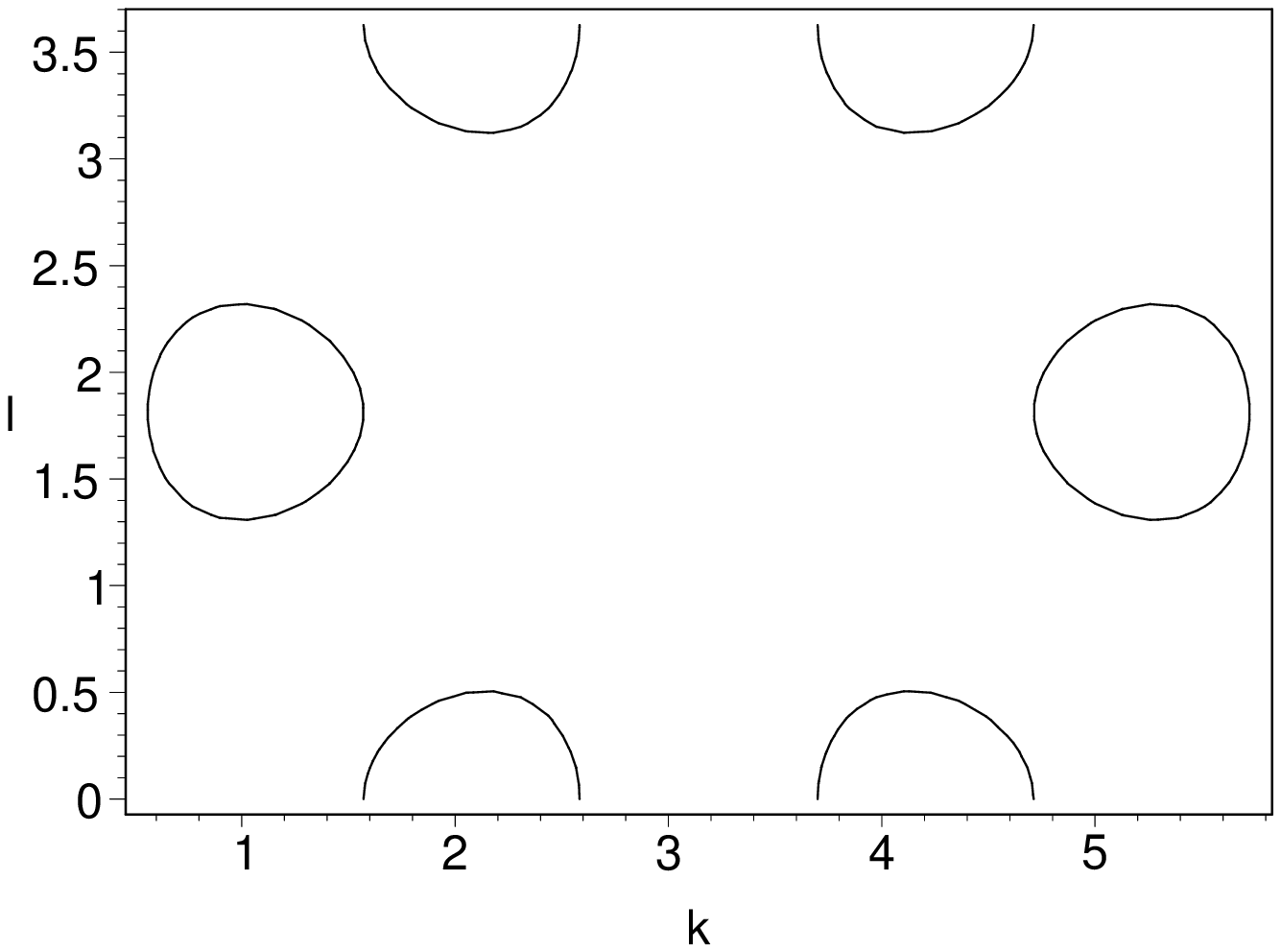}}
\put(-250,130){\footnotesize $\mcal{D}_1$}
\put(-200,44){\footnotesize $\mcal{D}_2$}
\put(-100,44){\footnotesize $\mcal{D}_3$}
\put(-48,130){\footnotesize $\mcal{D}_4$}
\put(-100,220){\footnotesize $\mcal{D}_5$}
\put(-200,220){\footnotesize $\mcal{D}_6$}
\caption{The domain $\mcal{D} = \mcal{D}_1 \cup \ldots \cup 
\mcal{D}_6$ in which the NLS equation \eref{hexrnls} is elliptic.}
\lbl{econt}
\end{center}
\end{figure}
\addtolength{\captionmargin}{+15pt}

Again, the hexagonal symmetry properties of the HETL are reflected clearly 
in the function $e(k,l)$, which has six maxima (at which $e(k,l)=36$), each
lying at the centre of one of the closed curves in \Fref{econt}.  The maxima 
of $e(k,l)$ coincide with the six maxima of $\w(k,l)$ shown in \Fref{wcont}, 
namely, at the wavevectors $\{\mb{k_1},\ldots,\mb{k_6}\}$ in $\mcal{T}^2$; 
$e(k,l)$ is minimised ($e(k,l)=-48$) at the six midpoints of the line 
segments which connect adjacent maxima.

\begin{figure}[ht]
\begin{center}
\resizebox{2.5in}{!}{\includegraphics{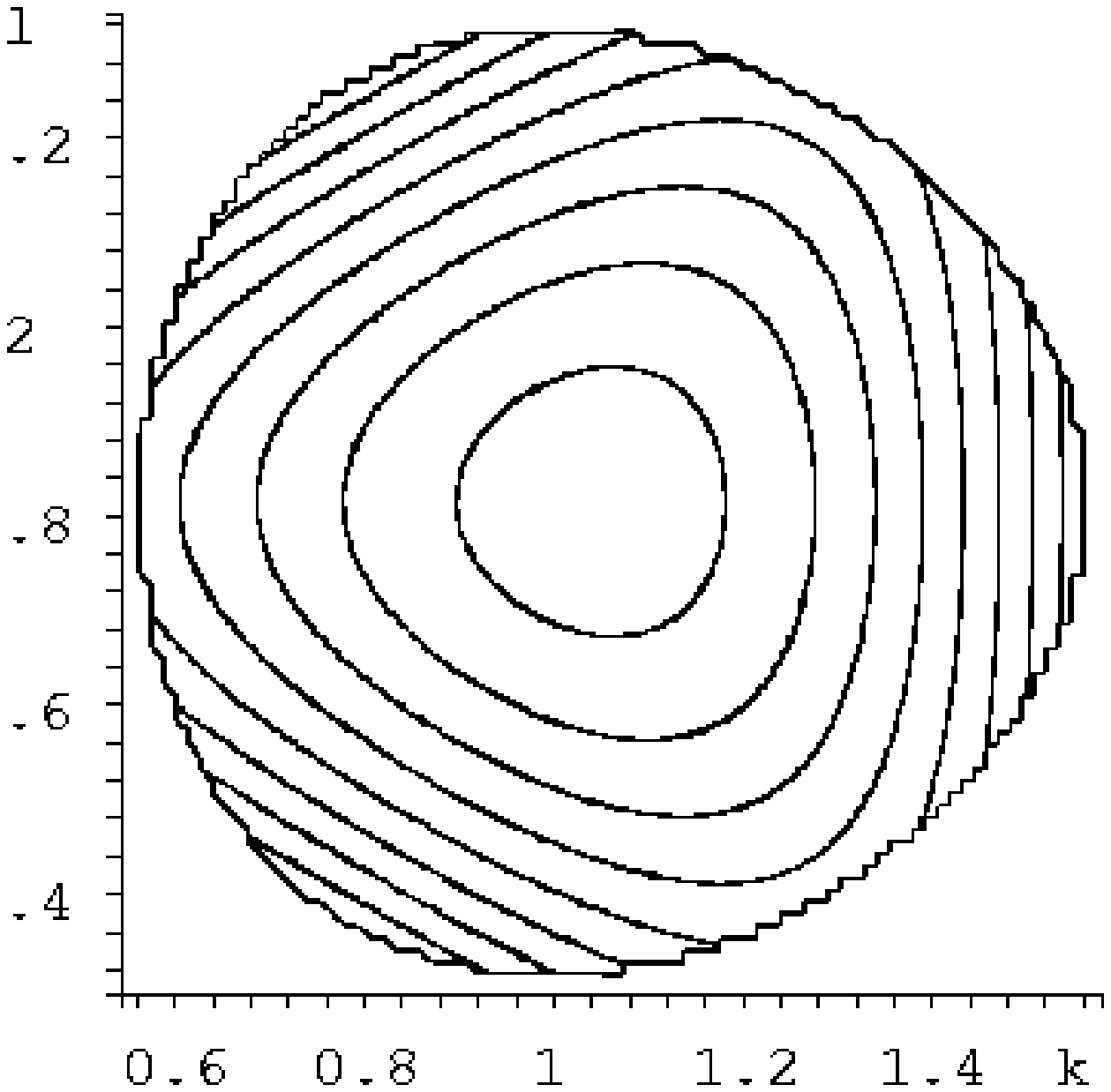}} \quad 
\resizebox{2.5in}{!}{\includegraphics{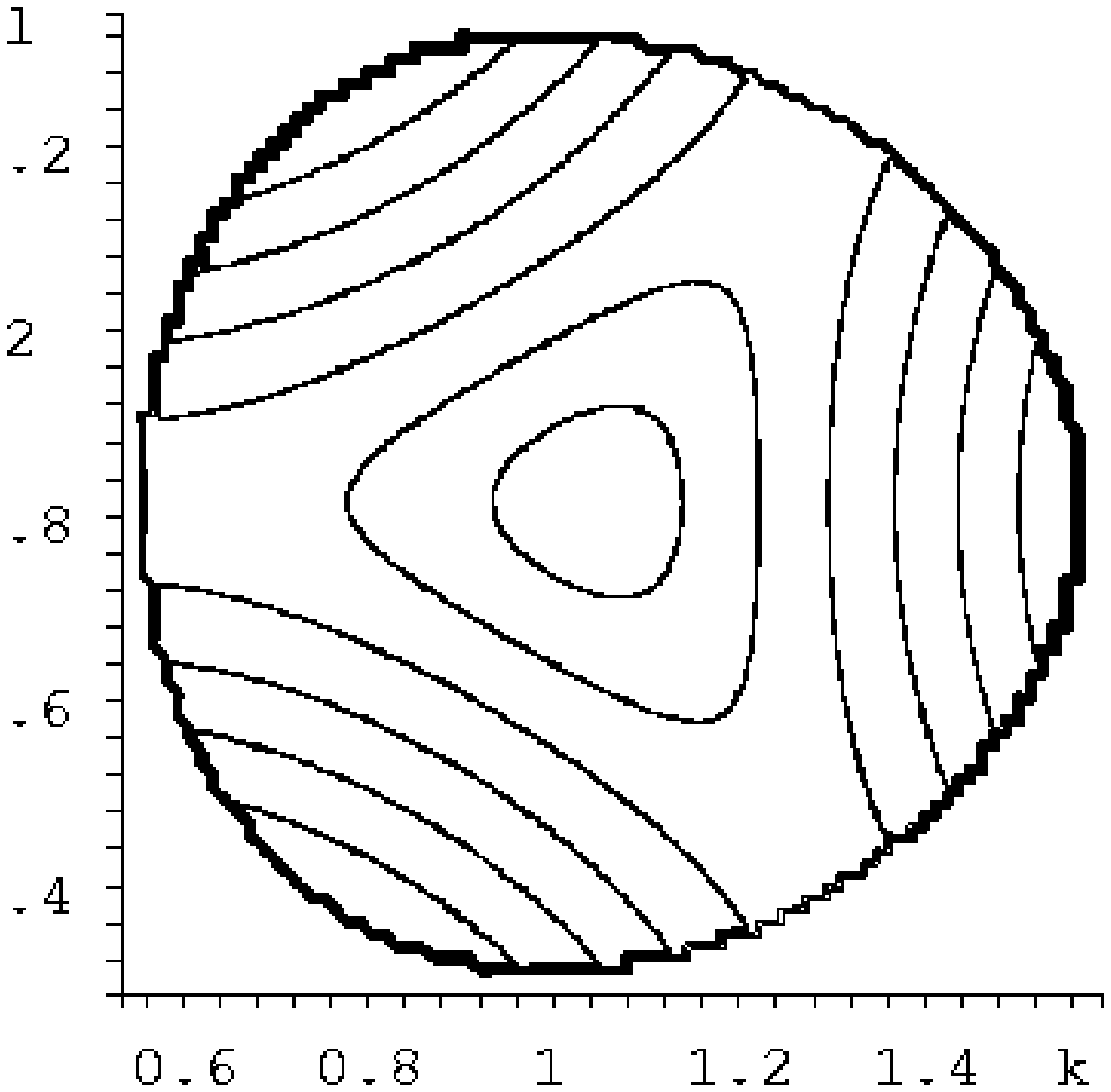}}
\end{center}
\caption{Contour plots of $\w(k,l)$ (left) and speed 
($\sqrt{u^2+v^2}$, right) for wavevectors $\mb{k}\in\mcal{D}_1$. 
Contours are for $2.82\leq\w\leq3$ in steps of $0.02$ and speeds from 
zero to 0.6 in steps of 0.1.  The central point corresponds to $\w=3$ and 
zero speed, as one considers wave vectors nearer to the edge of 
$\mcal{D}_1$, the frequency $\w$ decreases and the speed increases. }
\lbl{wspcont}
\end{figure}

Figure \ref{wspcont} shows a contour plot for the frequency $\w$ in 
$\mcal{D}_1$: the ellipticity constraint only permits breathers with a 
relatively high frequency, that is, with $\w>2.82$.  This constraint in turn 
implies that not all breather envelope velocities are attainable, only breathers 
with speeds upto about $0.7$ lattice sites per second are permitted; the plot 
on the right of Figure \ref{wspcont} shows that only breathers with speeds 
upto about $0.3$ sites per second can move in any direction. There is then an 
intermediate range of speeds, between $0.3$ and $0.7$ sites per second where 
breathers can only move in certain directions, these correspond to the lattice 
directions. The larger the speed, the more restricted is the direction of motion.

\subsubsection{Breather energy. } \lbl{hemv}

To calculate the leading-order energy $\elo$ of moving breathers in 
lattices with a symmetric potential we use \eref{hexlo} which we write as
\begin{eqn} \lbl{hxs2}
Q_{m,n}(t) \sim 2 \eee \alpha \cos \Phi \ \sech(\beta r),
\end{eqn}
where $\Phi = km + lhn + \om t$ is the phase of the carrier wave,
$(k,l) \in \mcal{D}$, $\om = \w + \eee^2\lambda$ is the breather
frequency including the first correction term,  ($\psi=km+lhn+\w t$ 
is only the leading-order expression).  $\ld$ parameterises the 
breather amplitude, $\alpha=\alpha(\lambda)$, $\beta=\beta(\lambda)$ 
and $r^2$ are as described in equations \eref{hexlo} and \eref{hexr}.

We now find expressions for the currents $I_{m,n}$, $J_{m,n}$ and
$K_{m,n}$: the current $I_{m,n}$ is obtained by substituting the 
expression for $Q_{m,n}$ \eref{hxs2} into \eref{hkiq} and integrating 
with respect to time.   Owing to the complexity of the expression for $r$ 
given by \eref{hexr}, the left-hand side of \eref{hkiq} can not be integrated 
with respect to time. However, the variable $r$ varies more slowly in time 
than $\Phi$. Hence,  integration by parts (using $\int f'(t) g(\eee t) \dd t = 
[f(t) g(\eee t)] - \eee \int f(t) g'(\eee t)\dd t$) gives, to leading-order, 
\begin{eqn} \lbl{himov} \fl
I_{m,n}\sim\frac{2\eee\alpha}{\w}\lft[1-\cos(2k)\rit]\sin\Phi\,\sech(\beta r)
- \frac{2\eee\alpha}{\w} \sin(2k) \cos \Phi \,\sech(\beta r),
\end{eqn}
where we have taken the constant of integration to be zero, and
$\om \sim \w$ to leading order.  Similarly, substituting for $Q_{m,n}$ in 
equations \eref{hkjq} and \eref{hkkq} and integrating, we find 
\begin{eqnarray} \fl
J_{m,n} & \sim & \frac{2\eee\alpha}{\w} \lft[ 1-\cos(k-lh) \rit] 
\sin \Phi\, \sech(\beta r) - 
\frac{2\eee\alpha}{\w}\sin(k-lh)\cos\Phi\,\sech(\beta r), \lbl{hjmov}\\[1ex] 
\fl  K_{m,n} & \sim & \frac{2\eee\alpha}{\w} \lft[ 1-\cos(k+lh) \rit] 
\sin \Phi\, \sech(\beta r)
- \frac{2\eee\alpha}{\w} \sin(k+lh) \cos \Phi \,\sech(\beta r). \lbl{hkmov}
\end{eqnarray}
Substituting these expressions into \eref{hentotlo}, we obtain 
\begin{eqnarray}
E_0 & \sim \sum_{m,n} & \frac{2\eee^2\alpha^2}{C_0} 
\cos^2 \Phi \ \sech^2(\beta r) \nonumber \\ & & + 
\frac{2L\eee^2\alpha^2}{\w^2}\sech^2(\beta r) \lft\{ \lft[ 
(1-\cos(2k))\sin\Phi - \sin(2k)\cos\Phi  \rit]^2 \right. \nonumber \\
& & + \lft[ (1-\cos(k-lh))\sin\Phi - \sin(k-lh)\cos\Phi \rit]^2 \nonumber \\ 
& & + \left. \lft[ (1-\cos(k+lh))\sin\Phi - \sin(k+lh)\cos\Phi \rit]^2 \rit\} .  
\lbl{hmvan} \end{eqnarray}
We replace the sum by an integral since the variables $Z=\eee (m-ut)$ 
and $W=\eee (hn-v t)$  vary slowly with $m$ and $n$.  To simplify the 
resulting integral, we approximate the terms $\cos^2\Phi$, $\sin^2\Phi$ 
and $\sin\Phi\cos\Phi$ by their average values of $\sfr{1}{2}$, $\sfr{1}{2}$ 
and $0$ respectively.  Hence \eref{hmvan} becomes 
\begin{eqn} \lbl{henmvint}
E_0 \sim \sum_{m,n} \frac{2\eee^2\alpha^2}{C_0} \sech^2(\beta r).
\end{eqn}

{}From the definition of $r$ (\ref{hexr}) we note the function 
$\sech(\beta r)$ is not in general radially symmetric in $m$, $n$.  Hence 
we work in $(\xi,\eta)$-space to facilitate evaluation of the double integral 
which approximates the double sum in (\ref{henmvint}).   Evaluating the 
Jacobian associated with the transformation from $(m,n)$ to $(\xi,\eta)$ 
coordinates, we find 
\begin{eqn}
E_0 \sim \frac{\alpha^2}{h^3C_0}\sqrt{4D_1D_2 - D_3^2} \int \!\!\! \int
\sech(\beta\sqrt{\xi^2+\eta^2}) \ \dd \xi \dd \eta. 
\lbl{hexemtri} \end{eqn}
Evaluating this and substituting for $\alpha$ and $\beta$ in terms of 
$D=3/2\w(k,l)$ and $B=3b\w(k,l)/2$ (see the appendix) we find 
\begin{eqn} \lbl{henar}
E_0 \sim \frac{4\pi\log 2(2\log 2 + 1)}
{3\, h^3 \, C_0 \, b \, \w^2(4\log 2 - 1)} \sqrt{4D_1D_2 - D_3^2}.
\end{eqn}
It is evident from \eref{henar} that the leading-order energy $E_0$ is 
independent of the breather amplitude, again confirming the existence 
of a minimum energy of moving breathers in the two-dimensional HETL 
with symmetric potential.  However, the threshold energy does depend 
upon the wavenumbers $k$ and $l$ hence moving breathers have a 
different threshold energy.   A plot of the expression \eref{henar} is 
shown in \Fref{henmov}.  We see that $\elo$, given by \eref{henar}, is 
strictly positive in the region of ellipticity $\mcal{D}$, and is maximised 
(attaining the same value) at each of the points corresponding to 
wavevectors $\{ \mb{k_1},\ldots, \mb{k_6} \}$, that is, at the points which
correspond to stationary breathers.  The threshold energy (\ref{henar}) 
decays to zero towards the boundary of the elliptic domain $\mcal{D}$ (see 
\Fref{econt}).   Hence, as for breathers in the square lattice, the energy 
threshold for moving breathers is {\em lower} than that for stationary 
breathers.  The threshold becomes arbitrarily small at the boundary of the 
domain of ellipticity.

\begin{figure}[htbp]
\begin{center}
\resizebox{3.5in}{!}{\includegraphics{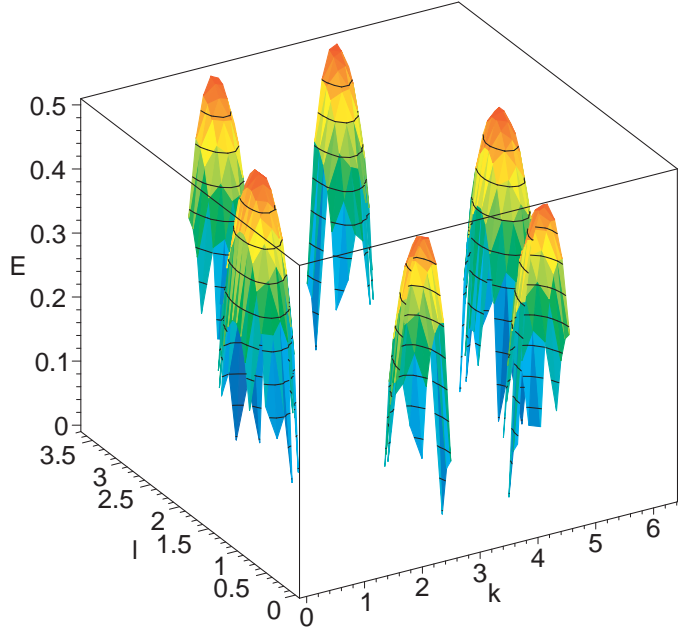}}
\caption{Plot of $E_0(k,l)$ for lattices with a symmetric potential.}
\lbl{henmov}
\end{center}
\end{figure}

\subsection{Lattices with an asymmetric potential} \lbl{hexasymm}

In this section, we consider the more general scenario where the potential 
$\Upsilon\!\ '(Q)$ is asymmetric.  In this case, the terms $a$ and $c$ in 
\eref{hqvexp} and \eref{heqq} are not both zero.   If \eref{hexnls} is to 
be reduced to an NLS equation in  $F$, both $G_0$ and $G_2$ must be 
found in terms of $F$.  Whilst $G_2$ is given by the simple algebraic 
equation \eref{hexg2}; in order to find $G_0$, the partial differential 
equation \eref{hexfo} must be solved.  We  assume that $G_0$ travels 
at the same velocity as $F$, that is, $G_0(X,Y,\tau,T) \equiv G_0(Z,W,T)$.  
Eliminating $G_{0\tau\tau}$,  \eref{hexfo} becomes
\begin{eqn} \lbl{hexg0}
(u^2 - 6)G_{0ZZ} + (v^2 - 2h^2)G_{0WW} + 2uvG_{0ZW} =
6a|F|_{ZZ}^2 + 2ah^2|F|_{WW}^2.
\end{eqn}
For general $k$ and $l$, it is difficult to solve for $G_0$ explicitly.  
However, for any of the wavevectors $\{\mb{k_1},\ldots,\mb{k_6}\}$, 
the velocities $u$ and $v$ become zero, and so \eref{hexg0} becomes 
$\nabla^2 G_0 = -a\nabla^2|F|^2$, where the operator $\nabla^2$, 
defined by $\nabla_{(Z,W)}^2 \equiv \clyd_Z^2 + \clyd_W^2$, is 
equivalent to $\nabla_{(X,Y)}^2 \equiv \clyd_X^2 + \clyd_Y^2$. It 
follows that $G_0 = -a|F|^2$ for each of these wavevectors.

Equations \eref{hexdisp}--\eref{hexfo} are the same whichever 
of the wavevectors $\{\mb{k_1},\ldots,\mb{k_6}\}$ is used.  Thus, 
\eref{hexdisp} gives $\w=3$, and  \eref{hexg2} implies $G_2 = aF^2/3$.  
Substituting these expressions for $G_0$ and $G_2$ in  \eref{hexnls} 
gives the following NLS equation for asymmetric potentials for any 
of the wavevectors $\{\mb{k_1},\ldots,\mb{k_6}\}$
\begin{eqn} \lbl{hnlsasy}
2\ii\w F_T + 3 \nabla^2 F + \w^2(3b-\mbox{$\frac{10}{3}$}a^2)|F|^2F=0,
\end{eqn}
The anomalous dispersion regime thus corresponds to $b>10a^2/9$.  
Soliton solutions are derived in the appendix;   \eref{hnlsasy} 
corresponds to $D = 1/2$ and $B =(9b-10a^2)/2$, and also 
$r^2 = X^2 + Y^2$.   Substituting the solution for $F$ into the lattice 
ansatz \eref{hexanz}, along with the known expressions for $G_0$ and 
$G_2$ gives the following second-order formula for stationary 
breathers in lattices with an asymmetric potential 
\begin{eqnarray} 
\fl Q_{m,n}(t) = 2\ \! \eee  \ \! \alpha\cos[km+lhn+(\w+\eee^2\lambda)t]\ 
\sech(\beta r) \nonumber \\ 
+ \mbox{$\frac{2}{3}$}\ \! a \ \! \eee^2 \ \! \alpha^2 \sech^2(\beta r)\!\lft\{ 
\cos[2km+2lhn+(2\w+2\eee^2\lambda)t] - 3 
\rit\} + \mcal{O}(\eee^3), \lbl{hexqasym} 
\end{eqnarray} 
where $\alpha$ and $\beta$ are determined in the appendix, 
$r=\sqrt{X^2+Y^2}$, and $X=\eee m$ and $Y=\eee h n$ in terms of the 
original discrete variables $m$ and $n$.

\subsubsection{Breather energy. } \lbl{hest}

We calculate the leading-order energy $\elo$ of stationary breathers in 
lattices with an asymmetric potential.  In this case, stationary breathers 
are given by \eref{hexqasym}, from which we take only the leading order term 
\begin{eqn} \lbl{hxa2}
Q_{m,n}(t) \sim 2 \eee \alpha \cos \Theta \ \sech(\beta r),
\end{eqn}
where $\Theta = km + lhn + \om t$ is the phase of the carrier wave 
with $(k,l)$ corresponding to one of $\{ \mb{k_1},\ldots, \mb{k_6}\}$, 
$\om = \w+\eee^2\lambda$ is the breather frequency, and $w=3$.

The currents $I_{m,n}$, $J_{m,n}$ and $K_{m,n}$ are obtained 
by substituting \eref{hxa2} into equations \eref{hkiq}--\eref{hkkq} 
and integrating with respect to time, taking the constant of 
integration to be zero.   Thus, to leading order, we find
\begin{eqnarray}
I_{m,n} & \sim & \frac{3\eee\alpha}{\w} \sin \Theta\, \sech(\beta r)
- \frac{\sqrt{3}\eee\alpha}{\w} \cos \Theta \,\sech(\beta r),\lbl{heni} \\[1ex]
J_{m,n} & \sim & \frac{3\eee\alpha}{\w} \sin \Theta\, \sech(\beta r)
+ \frac{\sqrt{3}\eee\alpha}{\w} \cos \Theta \,\sech(\beta r), \lbl{henj} \\[1ex]
K_{m,n} & \sim & \frac{3\eee\alpha}{\w} \sin \Theta\, \sech(\beta r)
+ \frac{\sqrt{3}\eee\alpha}{\w} \cos \Theta \,\sech(\beta r).  \lbl{henk}
\end{eqnarray}
Substituting these expressions into  \eref{hentotlo} 
gives the leading-order energy 
\begin{eqnarray} \fl
E_0 \sim \sum_{m,n} \frac{2\eee^2\alpha^2}{C_0} 
\cos^2 \Theta \ \sech^2(\beta r) \nonumber \\
+  \frac{Lh^2}{2\w^2}\,\eee^2\alpha^2  \lft[  3h^2\sin^2\Theta + 
2h\sin\Theta\cos\Theta + 3\cos^2\Theta  \rit]  \sech^2(\beta r)  . 
\lbl{henas}
\end{eqnarray}
We approximate the term in square brackets by taking the average
values of $\cos^2 \Theta$, $\sin^2\Theta$, $\sin\Theta\cos\Theta$ 
as \mbox{$\frac{1}{2}$}, \mbox{$\frac{1}{2}$} and $0$ respectively.

We replace the double sum by an integral giving
\begin{eqn} \lbl{henint}
E_0 \sim  \frac{2\alpha^2}{hC_0} \int \!\!\! \int
\sech^2(\beta \sqrt{X^2+Y^2}) \ \dd X \dd Y.
\end{eqn}
This can be evaluated to 
\begin{eqn} \lbl{henex}
E_0 \sim \frac{4\pi\log 2}{h C_0} \frac{\alpha^2}{\beta^2}
= \frac{8 \pi \log 2(2\log 2+1)}{C_0h(9b-10a^2)(4\log 2-1)} . 
\end{eqn}
Again, this estimate for energy is independent of the breather
amplitude $\lambda$, demonstrating the energy threshold properties
of the two-dimensional HETL; namely, the activation energy
required to create a breather in the HETL is an $\mcal{O}(1)$
quantity, irrespective of its amplitude ($2\eee\alpha\ll 1$).

\section{Higher-order asymptotic analysis}  \lbl{hex5}

The cubic NLS equation exhibits blow-up, which cannot occur in a discrete 
system since energy is conserved, and even if all the energy were localised 
at a single site, the amplitude would still be finite.   The two-dimensional 
cubic NLS equation description of the breather envelope is lacking.   
Higher-order dispersive and nonlinear effects play an important role in the 
dynamics of such discrete systems and therefore must be incorporated.  In 
this section we extend our analysis of the lattice equations \eref{heqq} to 
fifth-order, and derive a generalised NLS equation which includes higher-order 
dispersive and nonlinear terms.  It then remains to determine whether this 
generalised NLS equation supports stable soliton solutions.

Due to the complexity of a fifth-order analysis of  a general asymmetric 
potential, we consider only lattices with a symmetric potential, that is, 
those for which $a=c=0$ in \eref{heqq}. Since no second or fourth harmonic
terms are generated by the nonlinearity, we use a much simpler ansatz, namely 
\begin{eqn} \lbl{h5anzgen}
Q_{m,n}(t) = \eee \ee^{\ii \psi} F(X,Y,\tau,T)
+ \eee^3 \ee^{3\ii \psi}H_3(X,Y,\tau,T) + \cdots + \mathrm{c.c.,}
\end{eqn}
where the phase $\psi = km+lhn+\w t$.  In this case, in addition to the 
equations \eref{hexdisp2}--\eref{velsuv}, \eref{hexrnls} and \eref{hexh3}, 
we also have \\ 
\\
$\mcal{O}(\eee^5 \ee^{\ii\psi})$:
\begin{eqnarray} \fl 
F_{TT} =&& \hspace*{-12mm}
\mbox{$\frac{1}{6}$} \lft[ 8\cos(2k) + \cos k \cos(lh) \rit] F_{XXXX} 
+ h^2\cos k \cos(lh)F_{XXYY}  \nonumber \\ &&\hspace*{-16mm}
+ \mbox{$\frac{1}{6}$} h^4 \cos k \cos(lh)F_{YYYY} 
- \mbox{$\frac{2}{3}$} h \sin k \sin(lh) F_{XXXY} 
- \mbox{$\frac{2}{3}$} h^3 \sin k \sin(lh)F_{XYYY} 
\nonumber\\ &&\hspace*{-16mm}
+ 6b \lft[ 2\cos(2k) + \cos k \cos(lh) \rit](|F|^2F)_{XX}
+ 6bh^2\cos k\cos(lh)(|F|^2F)_{YY} \nonumber \\ && \hspace*{-16mm}
-12b\sin k \sin(lh) \lft[  2hFF_X\tbar{F}_Y  + 2hFF_Y \tbar{F}_X
+ 2h\tbar{F}F_XF_Y + hF^2\tbar{F}_{XY} + 2hF\tbar{F}F_{XY} \rit] 
\nonumber \\  && \hspace*{-16mm} 
- 3b\w^2\tbar{F}^2H_3 - 10\w^2 d |F|^4F . 
\lbl{hfogen} \end{eqnarray}
We simplify this by restricting attention to stationary breathers.
Accordingly, only one extra timescale, $T=\eee^2 t$, is required, and we 
fix the wavenumbers $k$ and $l$ to correspond to one of the wavevectors 
$\{ \mb{k_1},\ldots, \mb{k_6} \}$.  Hence we obtain the equations: \\
\\
$\mcal{O}(\eee^3 \ee^{\ii\psi})$: \\
\begin{eqn} \lbl{h5nls}
2\ii \w F_T + 3\nabla^2 F + 3b\w^2 |F|^2F = 0,
\end{eqn}
$\mcal{O}(\eee^5 \ee^{\ii\psi})$: \\
\begin{eqn} \lbl{h550}
F_{TT} = -\mbox{$\frac{3}{4}$} \nabla^4F - 9b\nabla^2(|F|^2F) -
10\w^2d|F|^4F , 
\end{eqn} 
in addition to $\w=3$ (from the $\mcal{O}(\eee \ee^{\ii\psi})$
equation) and $H_3=0$ (from $\mcal{O}(\eee^3 \ee^{3\ii\psi})$). 
A consequence of the hexagonal symmetry of the HETL is that all 
differentials on the right-hand side of \eref{h550} are isotropic.

In order to obtain a generalised NLS equation, we combine the
higher-order equation \eref{h550} with the cubic two-dimensional
NLS equation \eref{h5nls}.  The $F_{TT}$ term on the
left-hand side of \eref{h550} is eliminated  by
differentiating \eref{h5nls} with respect to $T$ and substituting
for $F_{TT}$ into \eref{h550}.  The resulting expression is 
\begin{eqnarray}
6 \ii F_T + 3\nabla^2 F + 27b|F|^2F + \frac{\eee^2}{2} \nabla^4 F
+ \frac{9\eee^2}{4}(40d-27b^2)|F|^4F & \nonumber \\ \qquad \qquad 
+ \frac{27b\eee^2}{4} \nabla^2(|F|^2F) - \frac{9b\eee^2}{2}|F|^2\nabla^2F
- \frac{9b\eee^2}{4} F^2\nabla^2\tbar{F} = 0 , & \lbl{hexgennls}
\end{eqnarray}
which is isotropic.

To the best of our knowledge, this perturbed form of the NLS equation 
has not been studied in the literature before; although it is similar to 
and slightly simpler than the corresponding equation derived for the 
square lattice \cite{buts06}.  It is also similar to the perturbed NLS 
equation considered by Davydova \etal \citep{dav03}, namely 
\begin{equation}
\ii F_T + D \nabla^2 F + B |F|^2 F + P \nabla^4 F + K |F|^4 F =0 , 
\lbl{davydova} 
\end{equation} 
which is known to have stable soliton solutions.  The anomalous 
dispersion case is $BD>0$, which in our equation (\ref{hexgennls}) 
corresponds to $b>0$;   Davydova's criteria for soliton existence is 
$PK>0$ which implies $40d > 27b^2$.  Hence, it is in this parameter 
regime that we seek breathers solutions of the HETL.  The numerical 
results presented in  \Sref{hexnumerics} show that long-lived breather 
solutions are supported by the two-dimensional hexagonal FPU lattice, 
suggesting that the additional perturbing terms in \eref{hexgennls}
do not destabilise the Townes soliton.
In fact, numerical simulations (presented in \cite{thesis} but not here) of the 
case $40d<27b^2$ show the breather mode to be long-lived, suggesting 
that the terms on the second line of (\ref{hexgennls}) are stabilising.

We have been unable to find a variational formulation of
\eref{hexgennls}, and are therefore unable to use the methods of 
Davydova {\em et al.}\ \cite{dav03} and Kuznetsov {\em et al.}\ 
\citep{kuz}.  Alternative possible methods include the modulation 
theory of Fibich and Papanicolaou \citep{fib98,fib99}.

\section{Numerical results}  \lbl{hexnumerics}

\subsection{Preliminaries} \lbl{hnumpre}

Using a fourth-order Runge-Kutta scheme, we solve the equations 
\begin{eqnarray}
\frac{\dd Q_{m,n}}{\dd t} & = & R_{m,n}, \nonumber \\[1ex]
\frac{\dd R_{m,n}}{\dd t} & = & (\delta^2_I + \delta^2_J +
\delta^2_K)
\lft[ Q_{m,n} + a Q_{m,n}^2 + b Q_{m,n}^3 + c Q_{m,n}^4 + d
Q_{m,n}^5 \rit], \lbl{heqqfo}
\end{eqnarray}
numerically.  Introducing the variable $R_{m,n}$ converts the 
system of second-order ordinary differential equations \eref{heqq} 
to an equivalent system of first-order differential equations.

We present the results of simulations for a range of parameter
values.  From \eref{velsuv}, the velocity of the envelope $(u,v)$ 
depends upon the wavevector $\mb{k}=[k,l]^T$, and hence we 
obtain moving breathers by choosing $(k,l) \in \mcal{D}$.  
In \Fref{ellcontnum}, we show the points in  $\mcal{D}_1 \subset  
\mcal{D}$ for which we solve the lattice equations \eref{heqqfo} 
numerically.  These points correspond to the wavevectors $\mb{k_1} 
= [\pi/3,\pi/h]^T$, $\mb{k_a} = [1.4,\pi/h]^T$, $\mb{k_b} = 
[0.79,1.7324]^T$ and $\mb{k_c} = [0.8,1.9987]^T$.  
Selecting $(k,l)$ too near to the boundary of $\mcal{D}$ results 
in a sharply elongated breather, which are difficult to simulate as 
they require a large domain.  Wavevectors close to any of 
$\{ \mb{k_1},\ldots, \mb{k_6} \}$, lead to breather modes with small 
speeds, a check of breather velocity would then require a long-time 
simulation.  In practice we have chosen wavenumbers which do not 
lead to a severely elongated breather envelope, and yet have velocities 
which result in observable displacements over reasonable times.   We 
present simulations of stationary breathers in systems with asymmetric 
potentials ($a,c \neq 0$) in \Sref{hnra} and, in Sections 
\ref{hnstsy}--\ref{hpsi130}, simulations of stationary and moving breathers 
in lattices with symmetric potentials (that is, $a=c=0$ in \eref{heqqfo}).

\begin{figure}[htbp]
\begin{center}
\resizebox{4in}{!}{\includegraphics{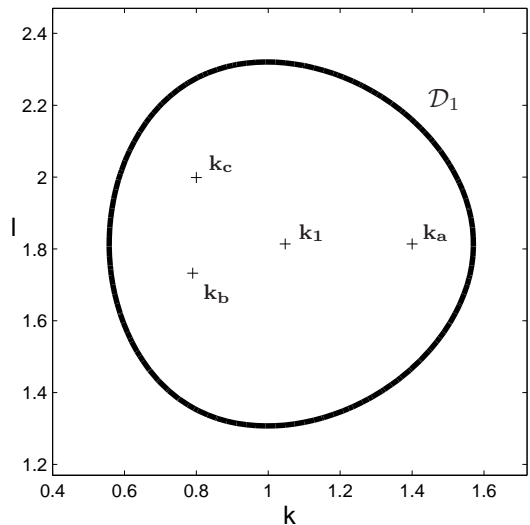}}
\put(-136,114){\scriptsize $\mb{k_1}$}
\put(-89,114){\scriptsize $\mb{k_a}$}
\put(-172,91){\scriptsize $\mb{k_b}$}
\put(-170,140){\scriptsize $\mb{k_c}$}
\put(-87,164){\fns $\mcal{D}_1$}
\caption{Wavevectors in $\mcal{D}_1 \subset \mcal{D}$ for 
which breathers are simulated.}
\lbl{ellcontnum}
\end{center}
\end{figure}

\subsection{Initial data and boundary conditions} \lbl{hbcics}

We generate initial data by using the analytic expressions for
breather solutions derived in \Sref{haa}. The formulae for $Q_{m,n}$ 
and $R_{m,n}$ are found in terms of the original discrete variables 
$m$ and $n$, and then shifted horizontally and vertically so that initially 
the breather lies at the centre of the lattice.   We impose periodic 
boundary conditions for the lattice in both horizontal and vertical 
directions, converting the two-dimensional arrangement illustrated 
in \Fref{hexetl} into a two-torus.  In long-time simulations, moving 
breathers which approach an edge of the lattice reappear from the 
opposite edge.  We select the site $(1,1)$ to lie at the bottom left-hand 
corner of the arrangement as illustrated in \Fref{hexbcs}.  Sites along 
the boundaries and corners are missing between one and four neighbours.  
We introduce fictitious sites along the boundaries and corners 
where necessary to effect periodic boundary conditions.

An illustration of a small finite lattice is shown in \Fref{hexbcs}.  
The dots represent capacitors located at lattice sites, and the 
lines represent some of the inter-connecting inductors.  From 
equations \eref{heqqfo},  the charge $Q_{m,n}$ stored on 
each capacitor depends upon the charge stored on the capacitors 
located at its six neighbouring sites,  two in each of the directions 
$\mb{e_i}$, $\mb{e_j}$ and $\mb{e_k}$.  
For the sake of clarity, the inductors connecting the capacitors at
the center of each hexagon to its six nearest neighbours are not
shown (see \Fref{hexetl}).  It is not necessary that the lattice
should be ``square,'' meaning that the lattice could comprise $M
\times N$ lattice sites, with $M \neq N$.  However, typically, we
consider $M=N$ as in \Fref{hexbcs}, since the numerical routines
are simpler to encode when the lattice \itc{is} square, and in  
the examples below, we consider lattices with $N \leq 50$.

\addtolength{\abovecaptionskip}{10pt}
\begin{figure}[htbp]
\begin{center}
\vspace{1.0cm}
\resizebox{2.5in}{!}{\includegraphics{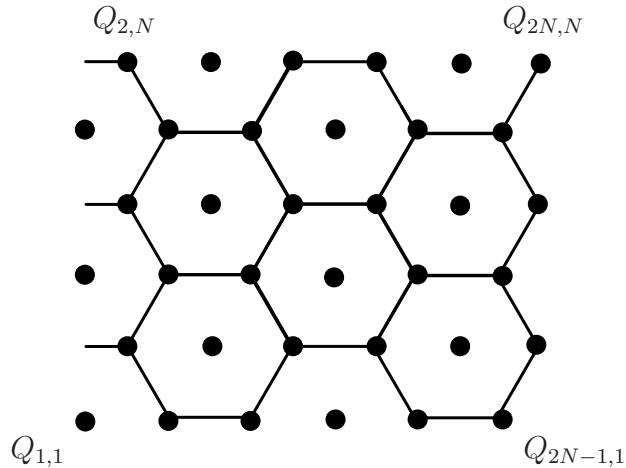}}
\put(-204,-10){${Q_{1,1}}$}
\put(-10,-10){${Q_{2N-1,1}}$}
\put(-173,153){${Q_{2,N}}$}
\put(-18,153){${Q_{2N,N}}$}
\caption{Periodic boundary conditions for the two-dimensional HETL.}
\lbl{hexbcs}
\end{center}
\end{figure}
\addtolength{\abovecaptionskip}{-10pt}

\subsection{Numerical computation of breather energy} \lbl{hnumen}

Since the HETL is lossless the total energy is conserved and 
can be used as a check of the accuracy of our numerical scheme.   
We compute the leading-order energy by expressing the summand 
$e_{m,n}^{(0)}$ \eref{hentotlo} in terms of the output variables of 
the numerical routine, namely $Q_{m,n}$ and $R_{m,n}$.  The first 
term of $e_{m,n}^{(0)}$ is simply dependent on $Q_{m,n}$.  It
remains to find the currents $I_{m,n}$, $J_{m,n}$ and $K_{m,n}$ in
terms of $Q_{m,n}$ and $R_{m,n}$.  The details of this calculation 
depend upon whether the interaction potential is symmetric or asymmetric.
As a check of the numerical scheme we verify that the sum is 
conserved and that it agrees with the asymptotic estimate \eref{henar}.

\subsubsection{Lattices with a symmetric potential. }

First, we find the current $I_{m,n}$ in terms of $Q_{m,n}$ and
$R_{m,n}$.  Differentiating the leading-order expression for the breather 
given by \eref{hxs2}, we have (retaining leading-order terms only)
\begin{eqn} \lbl{hnem}
\dot{Q}_{m,n} = R_{m,n} \sim -2\eee\alpha \w \sin \Phi \ \sech (\beta r),
\end{eqn}
where $\alpha$, $\beta$ and $r^2$ are as defined in \Sref{hexsym}.
Comparing the analytic expression \eref{himov} for the current
$I_{m,n}$ with the expressions for $Q_{m,n}$ \eref{hxs2} and
$R_{m,n}$ \eref{hnem}, we find
\begin{eqn} \lbl{hniqr}
I_{m,n} = \frac{\cos(2k)-1}{\w^2} R_{m,n} - \frac{\sin(2k)}{\w}Q_{m,n}.
\end{eqn}
Similarly, comparing the expressions \eref{hjmov} and \eref{hkmov}
for the currents $J_{m,n}$ and $K_{m,n}$ respectively with
equations \eref{hxs2} and \eref{hnem}, we find 
\begin{eqnarray}
J_{m,n} & = & \frac{\cos(k-lh)-1}{\w^2} R_{m,n} -
\frac{\sin(k-lh)}{\w}Q_{m,n},
\lbl{hnjqr} \\[1ex]
K_{m,n} & = & \frac{\cos(k+lh)-1}{\w^2} R_{m,n} -
\frac{\sin(k+lh)}{\w}Q_{m,n}. \lbl{hnkqr}
\end{eqnarray}
Substituting these expressions into \eref{hentotlo}, gives an expression 
for $E_0$ which does not simplify, so we do not reproduce it here.

\subsubsection{Lattices with an asymmetric potential. }

Differentiating the leading-order expression \eref{hxa2} for the
charge $Q_{m,n}$ in lattices with an asymmetric potential gives
\begin{eqn} \lbl{hxa2dot}
\dot{Q}_{m,n} = R_{m,n} \sim 
-2\eee\alpha \w \sin \Theta \ \sech (\beta r) , 
\end{eqn}
where $\Theta=\pi m/3 +\pi n + \om t$ with $\om=3$.  Using \eref{hxa2} 
and \eref{hxa2dot}, the current $I_{m,n}$  given by \eref{heni} can be 
expressed in terms of $Q_{m,n}$ and $R_{m,n}$ as
\begin{eqn} \lbl{hiqqd}
I_{m,n} \sim -\frac{\sqrt{3}}{2\w} \lft[ \frac{\sqrt{3}R_{m,n}}{\w} + 
Q_{m,n} \rit]\!.
\end{eqn}
Similarly, from  \eref{henj}--\eref{henk}, we have
\begin{eqn} \lbl{hjkqqd}
J_{m,n}  = K_{m,n} \sim -\frac{\sqrt{3}}{2\w} 
\lft[ \frac{\sqrt{3}R_{m,n}}{\w} - Q_{m,n} \rit]\!.
\end{eqn}
These are equivalent to substituting $k=\pi/3$ and $l=\pi/h$ into 
(\ref{hniqr})--(\ref{hnkqr}).  Inserting the expressions for $I_{m,n}$, 
$J_{m,n}$ and $K_{m,n}$ (\eref{hiqqd}--\eref{hjkqqd}) into \eref{hentotlo} yields 
\begin{eqn} \lbl{hentotqqd}
E_0 = \sum_{m,n} \frac{1}{72 C_0} \lft[ 
45Q_{m,n}^2 + 3R_{m,n}^2 - 2\sqrt{3}Q_{m,n}R_{m,n} \rit]\! . 
\end{eqn}

\subsubsection{Effective breather width. }

Numerical computation of the energy as described above allows us to 
check that the total lattice energy is conserved, but does not indicate 
whether a breather changes shape over time.  To remedy this, we define 
breather widths in the $m$ and $n$ directions, the sum of which we 
denote $\mcal{W}_{\mathrm{br}}$, where
\begin{eqn} \lbl{hbrwdth}
\mcal{W}_{\mathrm{br}}^2 = \frac{r_{20}}{\elo} + \frac{r_{02}}{\elo}
- \lft( \frac{r_{10}}{\elo} \rit)^2 -  \lft( \frac{r_{01}}{\elo} \rit)^2,
\end{eqn}
and $r_{10}$, $r_{01}$, $r_{20}$ and $r_{02}$ are defined by
\addtolength{\arraycolsep}{0.5cm}
\begin{displaymath}
\begin{array}{c c}
r_{10}=\sum_{m,n} m e_{m,n},\quad&r_{20}=\sum_{m,n}m^2 e_{m,n},\\[8pt]
r_{01}=\sum_{m,n} hn e_{m,n},&r_{02} = \sum_{m,n} h^2n^2 e_{m,n}. 
\end{array}
\end{displaymath}
\addtolength{\arraycolsep}{-0.5cm}
The variation in $\wbr$ over time gives a measure of the distortion 
suffered by a breather.

 \subsection{Stationary breather in a lattice with symmetric potential}
\lbl{hnstsy}

We investigate stationary breather solutions in the anomalous dispersion
regime which corresponds to $b>0$ (see \eref{hexgennls}).  We set 
$k=\pi/3$ and $l=\pi/h$, corresponding to $\mb{k_1}$ in \Fref{ellcontnum}, 
so that the velocities $u$ and $v$ are zero.   Davydova's result 
implies that we expect to find stable soliton solutions when $PK>0$.  
{}From \eref{davydova}, $P=\eee^2/2$, hence this inequality implies 
$40d > 27b^2$.  Hence we choose $b=d=1$ as well as $N=30$, $\eee=0.2$ 
and $\lambda=1$.  Using the technique outlined in the appendix we 
calculate $\alpha=1.0212$, $\beta=1.8670$.  The breather frequency is 
$\w+\eee^2\lambda=3.040$, and therefore the period of oscillation is $T=2.0668$.

\begin{figure}[htbp]
\centering 
\subfigure[Profile at $t=0$.]{
\includegraphics[width=.45\textwidth]{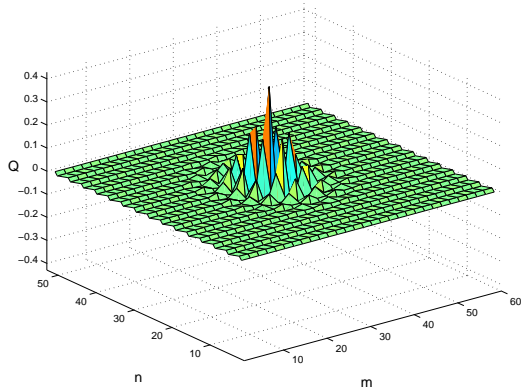}}
\hspace{.3in} 
\subfigure[Plot of $e_{m,n}$, $E_0=0.7606$.]{
\includegraphics[width=.45\textwidth]{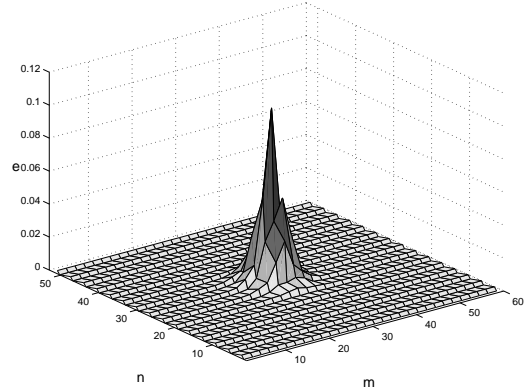}}\\
\vspace{.3in} 
\subfigure[Profile at $t=30T=62.82$.]{
\includegraphics[width=.45\textwidth]{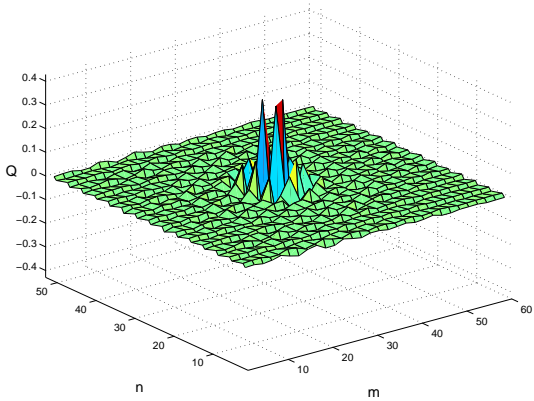}}
\hspace{.3in} 
\subfigure[Plot of $e_{m,n}$, $E_0=0.7087$.]{
\includegraphics[width=.45\textwidth]{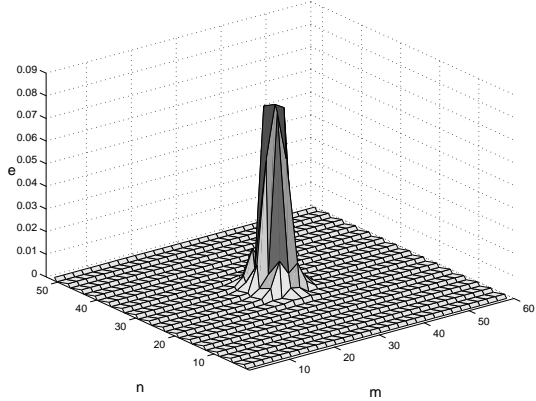}}
\caption{Stationary breather in a lattice with symmetric potential, 
see Section \ref{hnstsy} for details.}
\label{simh1}
\end{figure}

As is the case for all our simulations, the initial profile of the breather 
is located at the centre of the lattice, as illustrated in  \Fref{simh1}\,(a). 
At $t=0$, the breather energy is $\elo=0.7606$, and $\wbr=3.74$.
The asymptotic estimate for $E_0$ given by \eref{henar} is 0.7523, 
which is only 1\% different from the numerically obtained value.
In \Fref{simh1}\,(c) we show the breather after 30 oscillations.  
Plots of the local energy $e_{m,n}$ at $t=0$
and $t=30T$ are presented in Figures \ref{simh1}\,(b) and
\ref{simh1}\,(d).  After thirty oscillations the breather has shed 
a small amount of energy, which is manifested as small amplitude 
radiation throughout the lattice.  Accordingly, the breather 
appears a little distorted in shape compared to its initial profile.  
In particular, at $t=30T$, we find $\wbr=4.21$. The energy of the 
breather at $t=30T$ is 0.7087, giving $\Delta E_0/E_0=-0.0682$,

\subsection{Breather moving along a lattice direction ($\Psi=0\degrees$)}
\lbl{hpsi0}

\begin{figure}[t]
\centering 
\subfigure[Profile at $t=0$, $E_0=0.5537$.]{
\includegraphics[width=.45\textwidth]{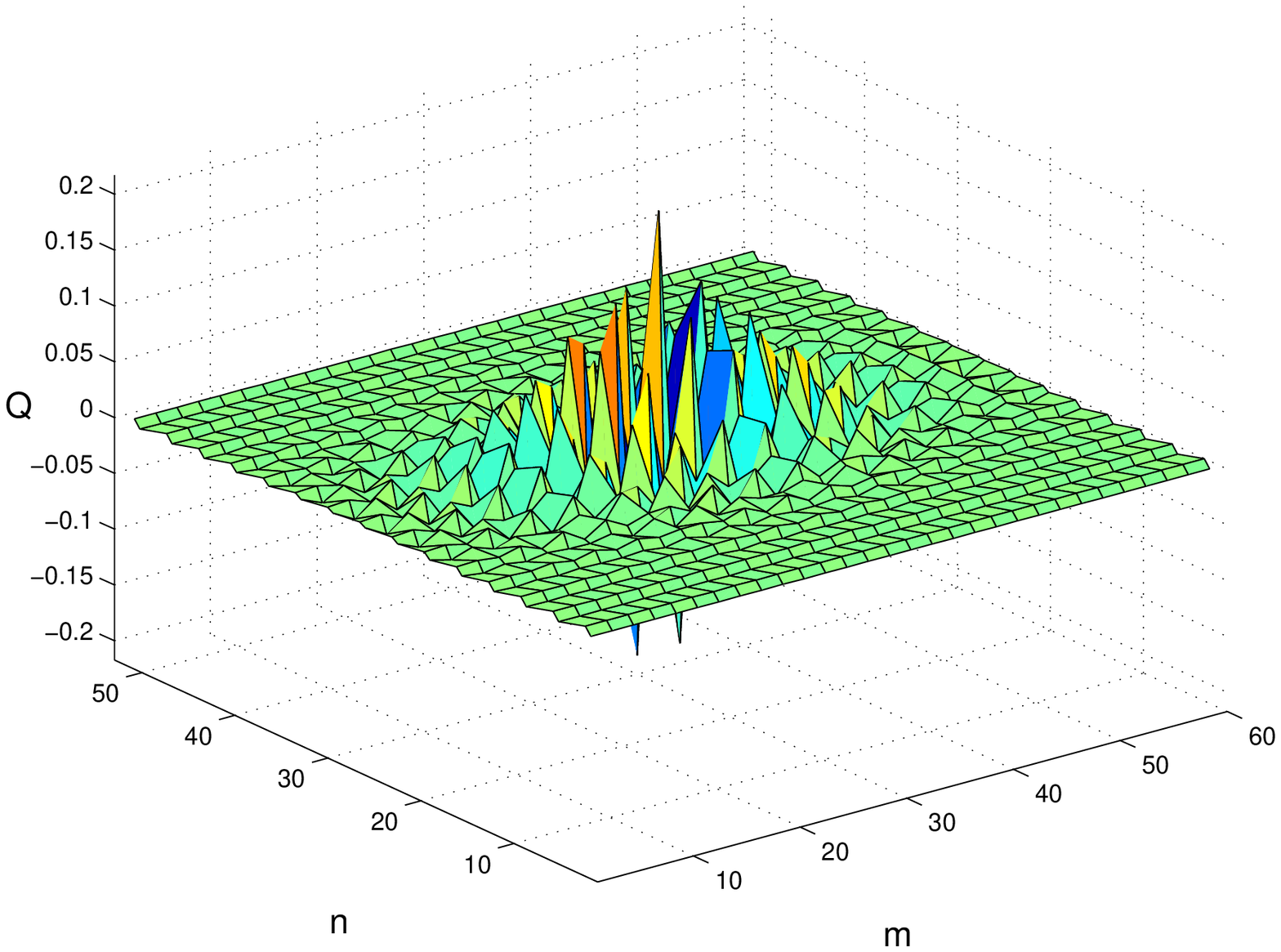}}
\hspace{.3in} 
\subfigure[Profile at $21.58T$, $E_0=0.5562$.]{
\includegraphics[width=.45\textwidth]{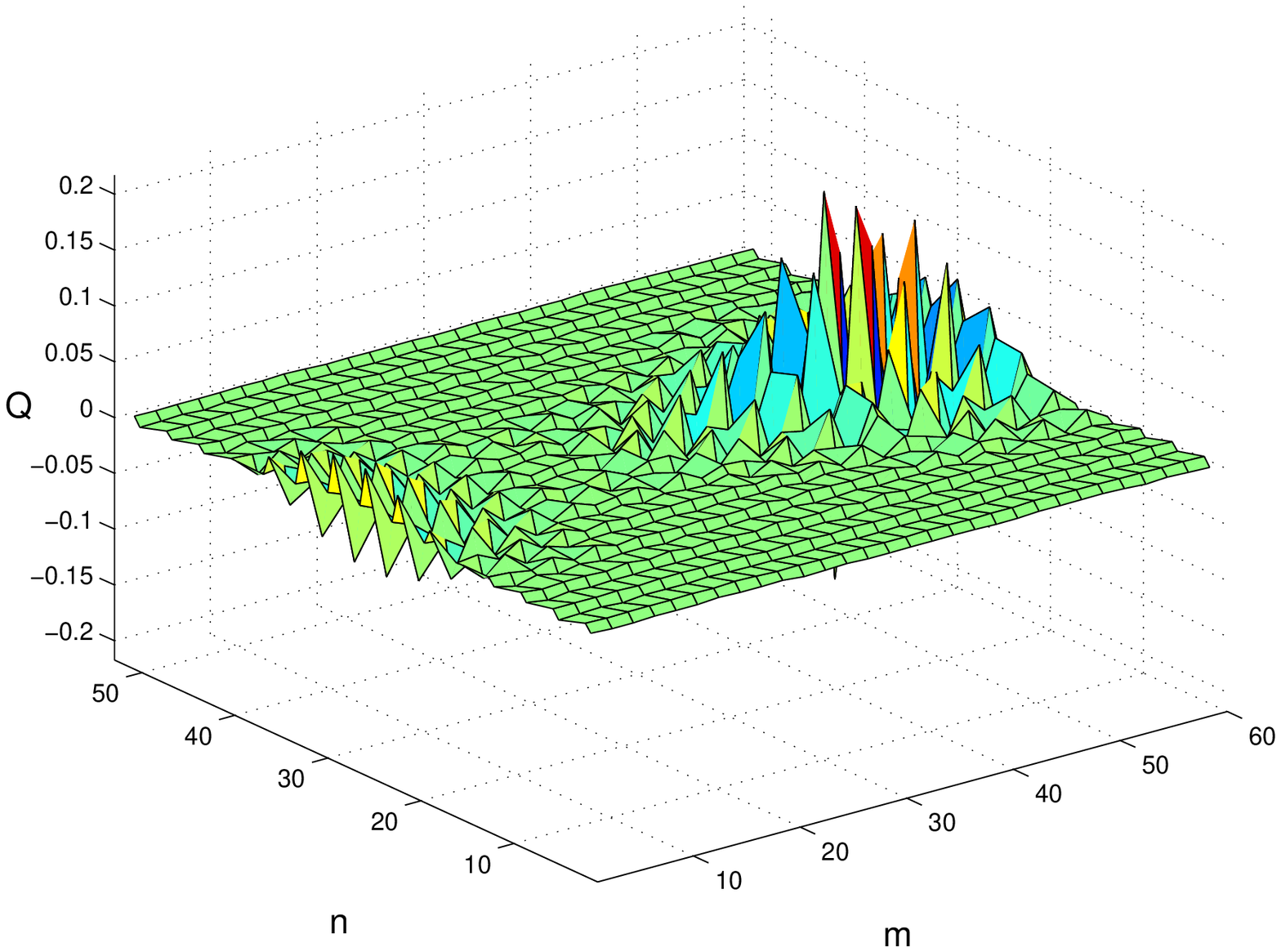}}\\
\vspace{.3in} 
\subfigure[Profile at $t=46.59T$, $E_0=0.5696$.]{
\includegraphics[width=.45\textwidth]{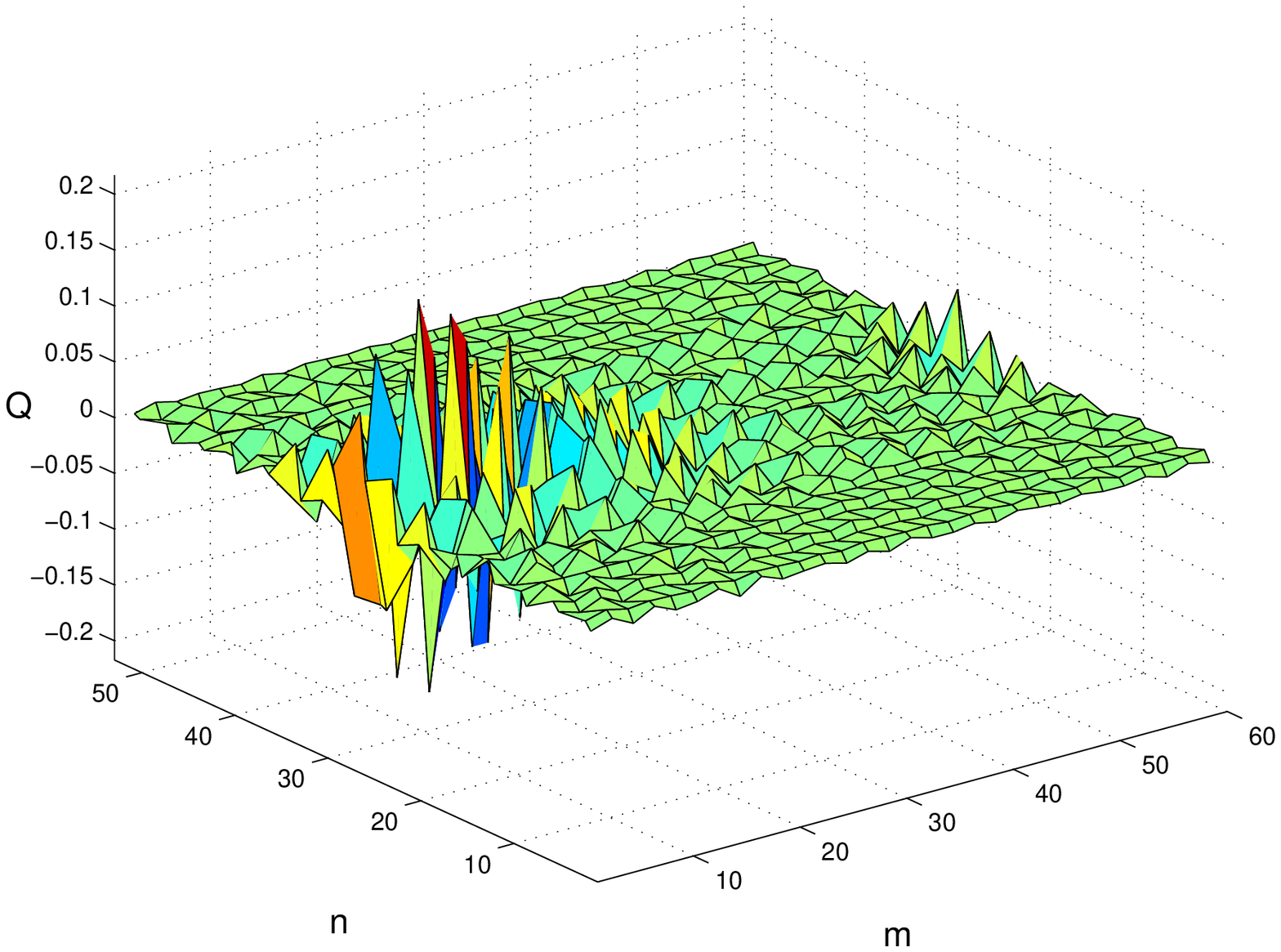}}
\hspace{.3in} 
\subfigure[Plot of $e_{m,n}$.]{
\includegraphics[width=.45\textwidth]{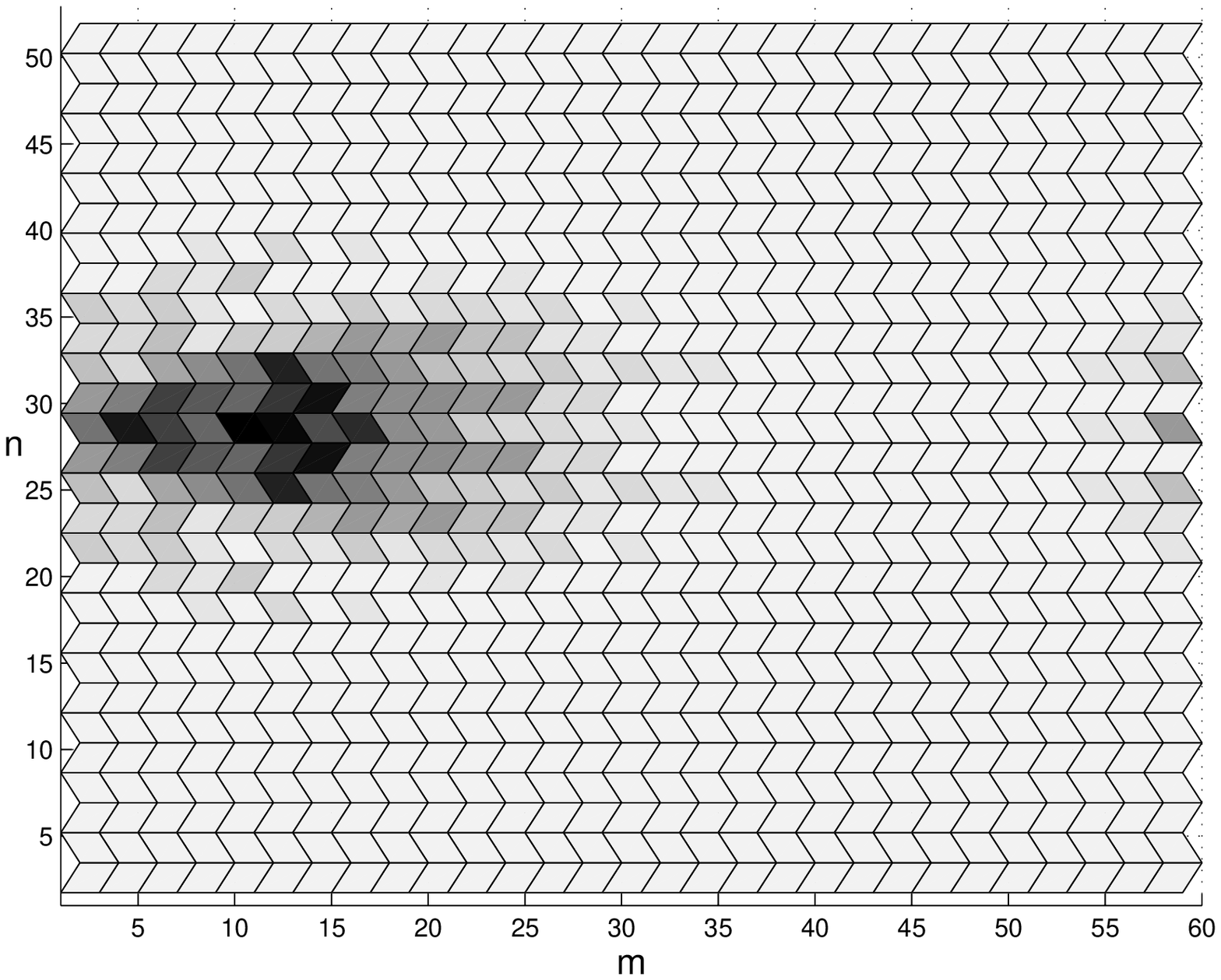}}
\caption{Breather moving along a lattice direction, $\Psi=0\degrees$, 
see Section \ref{hpsi0} for details.}
\label{simh2}
\end{figure}

In \Fref{simh2}, we show a simulation of a breather moving along a
lattice direction parallel to the $m$-axis, that is, $\Psi=0\degrees$.  We 
have chosen $\mb{k} = \mb{k_a} = [1.4,\pi/h]^T$ so that $u=0.4445$ 
and $v=0$ and thus $\Psi=\tan^{-1}(v/u) = 0\degrees$.  The breather 
frequency is $\w=2.9265$, and hence the period $T=2.1470$. Following 
the calculation outlined in the appendix, the amplitude and width 
parameters are $\alpha=1.0339$ and $\beta=1.8440$; the remaining 
parameters being $b=1$, $d=1$, $N=30$, $\eee=0.1$ and $\lambda=1$.

The initial profile of the breather is shown in \Fref{simh2}\,(a), and at 
this time, the calculated energy is $E_0=0.5537$, whilst the asymptotic 
estimate \eref{henar} is $E_0=0.5522$.  It may be observed that the 
breather is not radially symmetric. In fact, it is slightly elongated in the 
direction parallel to the $m$-axis, that is, parallel to the direction of 
motion. This is because the point corresponding to the wavevector 
$\mb{k_a}$ is near the boundary of the region of ellipticity $\mcal{D}_1$ 
as illustrated in \Fref{econt}.

\Fref{simh2}\,(b) shows the breather at the later time of  $t=21.58T=46.3513$, 
at which time we find $\elo=0.5562$.  By this time, the breather has reached 
the right-hand edge of the lattice and, owing to periodic boundary conditions, 
it reemerges from the left-hand side as shown in \Fref{simh2}\,(c). The 
breather remains localised, without spreading, though it leaves behind a 
small amount of energy in its path (visible in Figures \ref{simh2}\,(b) and 
\ref{simh2}\,(c)).   The computed energy changes much less than for the 
previous simulation (here, $\Delta E_0/E_0=0.0287$).

The velocity of the breather can be measured from a plot of the energy 
$e_{m,n}$, which is shown in \Fref{simh2}\,(d), viewed from directly 
above the plane of the lattice.  From this plot, we note that  the 
breather has travelled 42 units at an average speed of 0.42 units per 
second, which is 5.5\% lower than our predicted speed of 0.4445 units 
per second.  This is consistent with a small amount of energy being 
shed as the system transforms from our approximated initial conditions 
into the precise shape of the breather.

\subsection{Breather moving at $\Psi=210\degrees$} \lbl{hpsi210}

\begin{figure}[t]
\centering 
\subfigure[Profile at $t=25$, $\elo = 0.6295$.]{
\includegraphics[width=.45\textwidth]{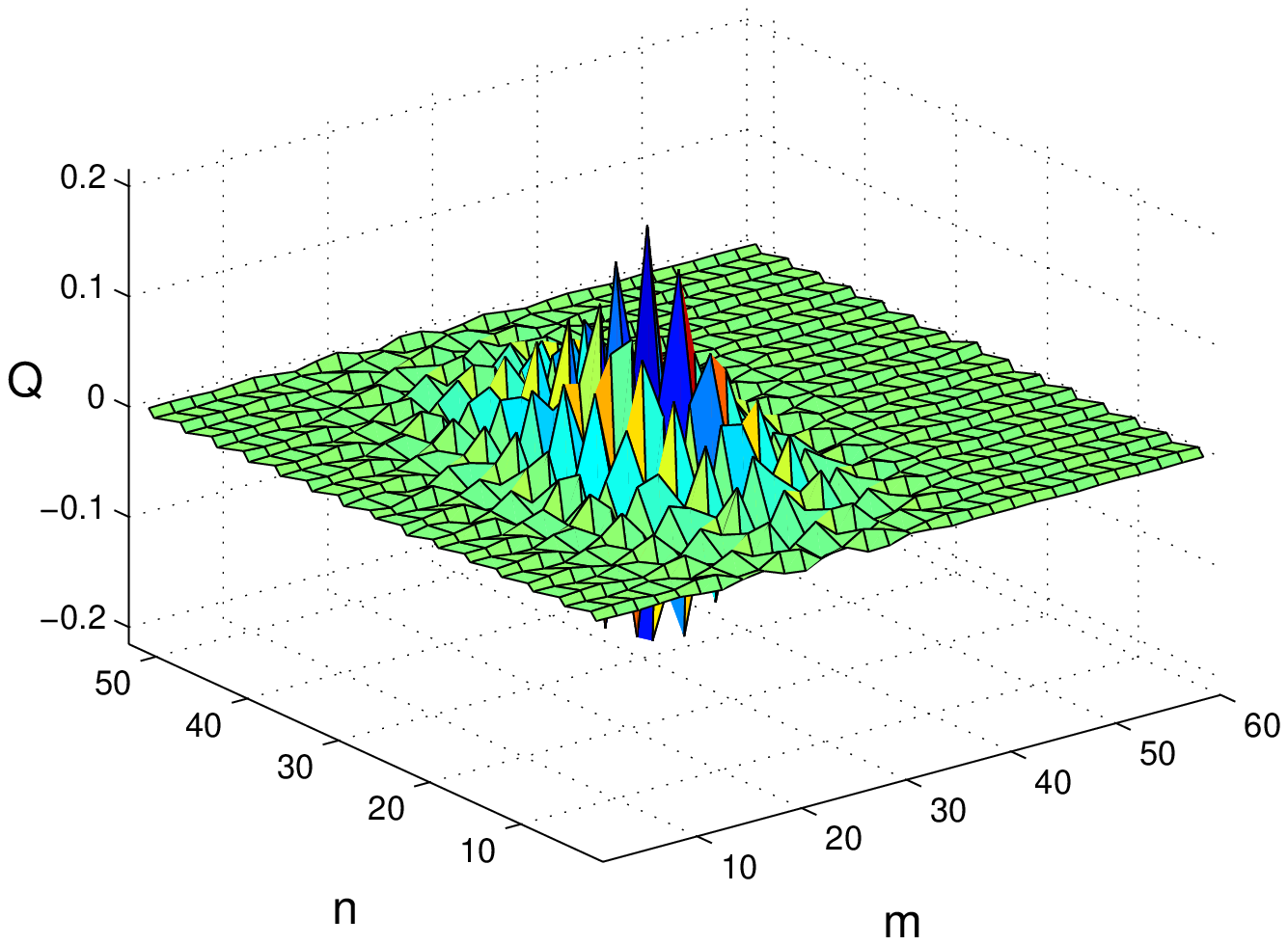}}
\hspace{.3in} 
\subfigure[Profile at $t=50$, $\elo = 0.6292$.]{
\includegraphics[width=.45\textwidth]{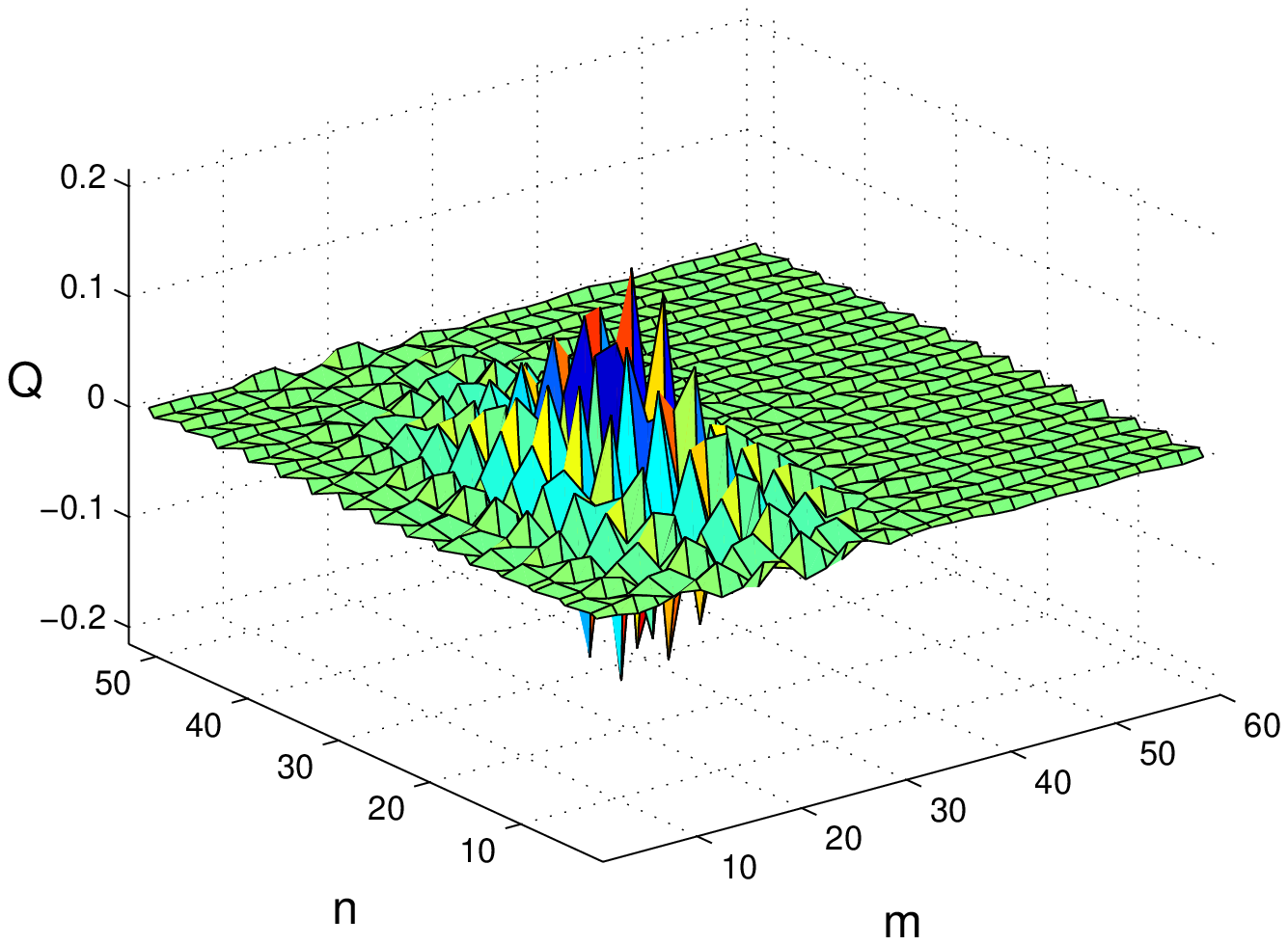}}\\
\vspace{.3in} 
\subfigure[Profile at $t=75$, $\elo = 0.6235$.]{
\includegraphics[width=.45\textwidth]{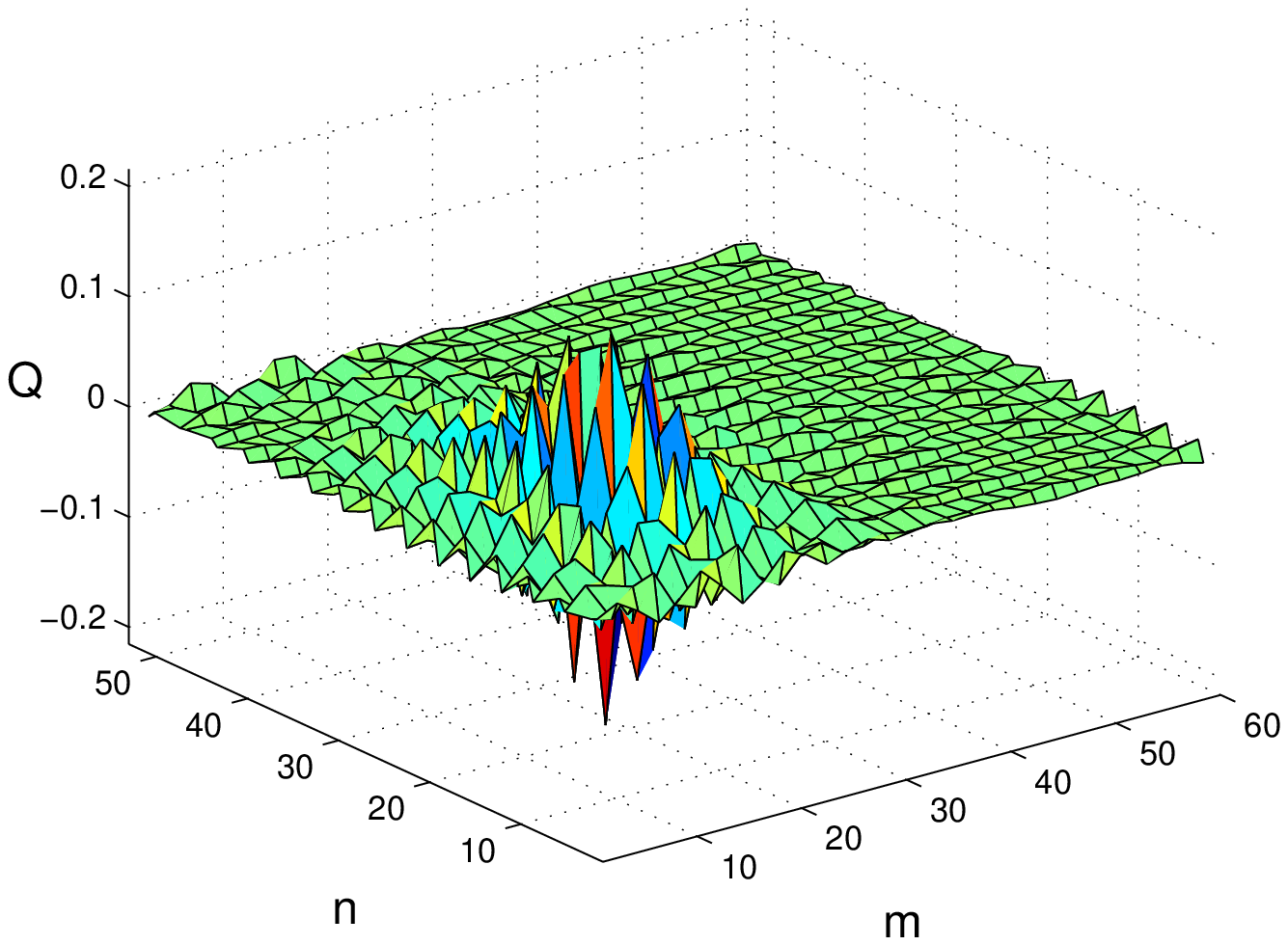}}
\hspace{.3in} 
\subfigure[Profile at $t=100$, $\elo = 0.6275$.]{
\includegraphics[width=.45\textwidth]{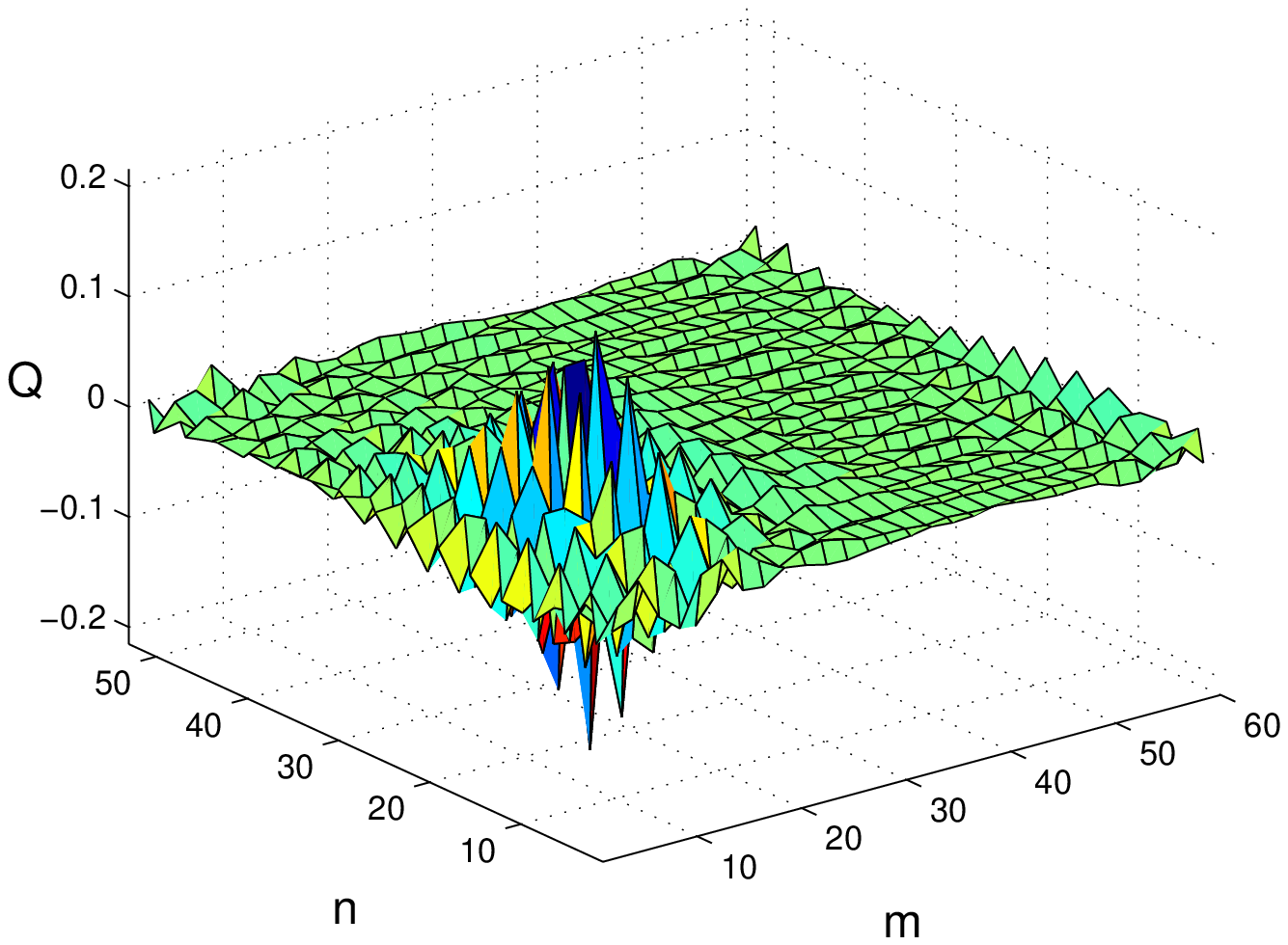}}
\caption{Breather moving at $\Psi=210\degrees$, 
see Section \ref{hpsi210} for details.}
\label{simh6}
\end{figure}

We now show that it is possible to simulate breathers moving in
directions other than a lattice direction (that is, $\Psi \neq 0\degrees$).  
We set $k=0.79$ and $l=1.7324$, which corresponds to $\mb{k_b}$ in
\Fref{ellcontnum}.  It may be verified that $u=-0.1999$ and
$v=-0.1154$ units per second, leading to $\Psi = 210\degrees$ as 
required.  Although not a lattice vector, this direction is an axis of 
symmetry of the lattice. For the wavevector $\mb{k_b}$, we have 
$\w=2.9675$ and
hence $T=2.1174$,  the remaining parameters being $b=1$, $d=1$,
$N=30$, $\eee=0.1$ and $\lambda=1$.  The variational parameters
$\alpha$ and $\beta$ are 1.0267 and 1.8568 respectively.  The
breather is shown at times $t=25$, 50, 75 and 100 seconds in
\Fref{simh6}.  Clearly, the breather does not deform significantly
as it travels, nor does it radiate much energy.  The initial energy is 
computed to be $E_0=0.6306$, and at $t=100$, the energy is 0.6275, 
a loss of 0.5\%.   The asymptotic estimate for the energy is 
$E_0=0.6184$ --- 2\% different from the numerically computed value.
The motion of the breather is charted in \Tref{tsimh6}, the final 
measured values for the velocities $u$ and $v$ give an average 
speed of 0.2152 units per second, 6.8\% below the predicted 
speed of 0.2308 units per second. The angle of travel is 
almost identical to the expected value of $\Psi=210\degrees$.

\setlength{\arrayrulewidth}{1pt}
\begin{table}
\begin{center}
\begin{tabular}[htbp]{| c | c | c | c | c | c | c |} \hline 
{\ssz{Time}} & {\ssz{Horizontal}} & {\ssz{Vertical}} & 
{\ssz{Average horizontal}} & {\ssz{Average vertical}} & {} & {} \\[-8pt]  
{\ssz{(s)}} & {\ssz{displacement}} & {\ssz{displacement}} & 
{\ssz{velocity (units s$^{-1}$)}} & {\ssz{velocity (units s$^{-1}$)}} & 
{\ssz{$\tan\Psi$}} & {\ssz{$\Psi$}} \\  \hline 
{\ssz{$25$}} & {\ssz{$-3.5$}}  & {\ssz{$-2$}}  & {\ssz{$-0.14$}}  & {\ssz{$-0.08$}}  & {\ssz{$0.5714$}} & {\ssz{$209.74\degrees$}} \\ \hline 
{\ssz{$50$}} & {\ssz{$-8$}}  & {\ssz{$-4.5$}}  & {\ssz{$-0.16$}}  & {\ssz{$-0.09$}}  & {\ssz{$0.5625$}} & {\ssz{$209.36\degrees$}} \\ \hline 
{\ssz{$75$}} & {\ssz{$-13$}}  & {\ssz{$-7.5$}}  & {\ssz{$-0.1733$}}  & {\ssz{$-0.1$}}  & {\ssz{$0.5770$}} & {\ssz{$209.99\degrees$}} \\ \hline 
{\ssz{$100$}} & {\ssz{$-18.5$}}  & {\ssz{$-11$}}  & {\ssz{$-0.185$}}  & {\ssz{$-0.11$}}  & {\ssz{$0.5946$}} & {\ssz{$210.74\degrees$}} \\ \hline
\end{tabular}
\end{center}
\caption{Summary of breather motion ($\Psi=210\degrees$).}
\label{tsimh6}
\end{table}

\subsection{Breather moving at $\Psi=130\degrees$} \lbl{hpsi130}

\begin{figure}[t]
\centering 
\subfigure[Profile at $t=30$, $\elo = 0.5961$.]{
\includegraphics[width=.45\textwidth]{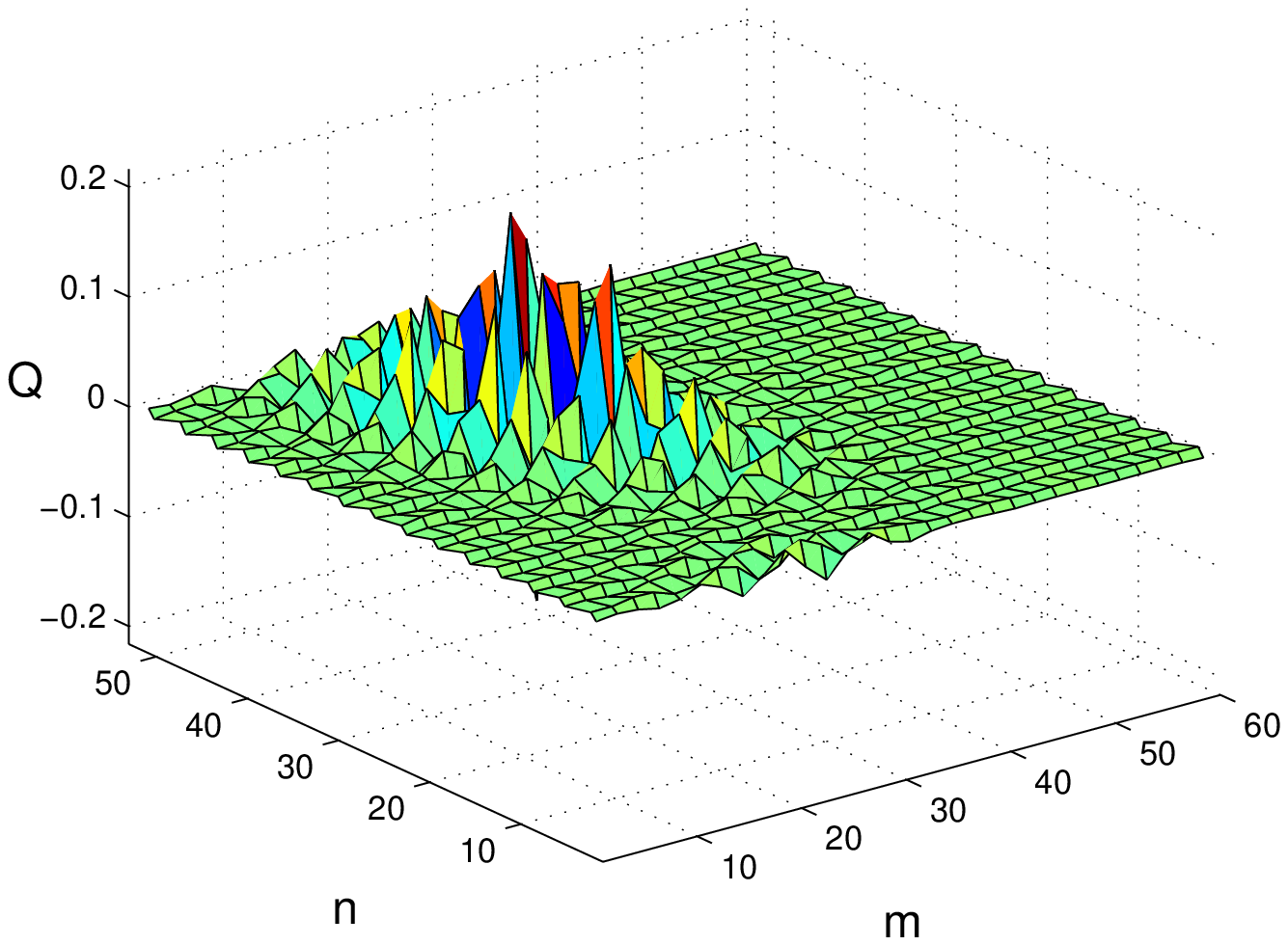}}
\hspace{.3in} 
\subfigure[Profile at $t=60$, $\elo = 0.5966$.]{
\includegraphics[width=.45\textwidth]{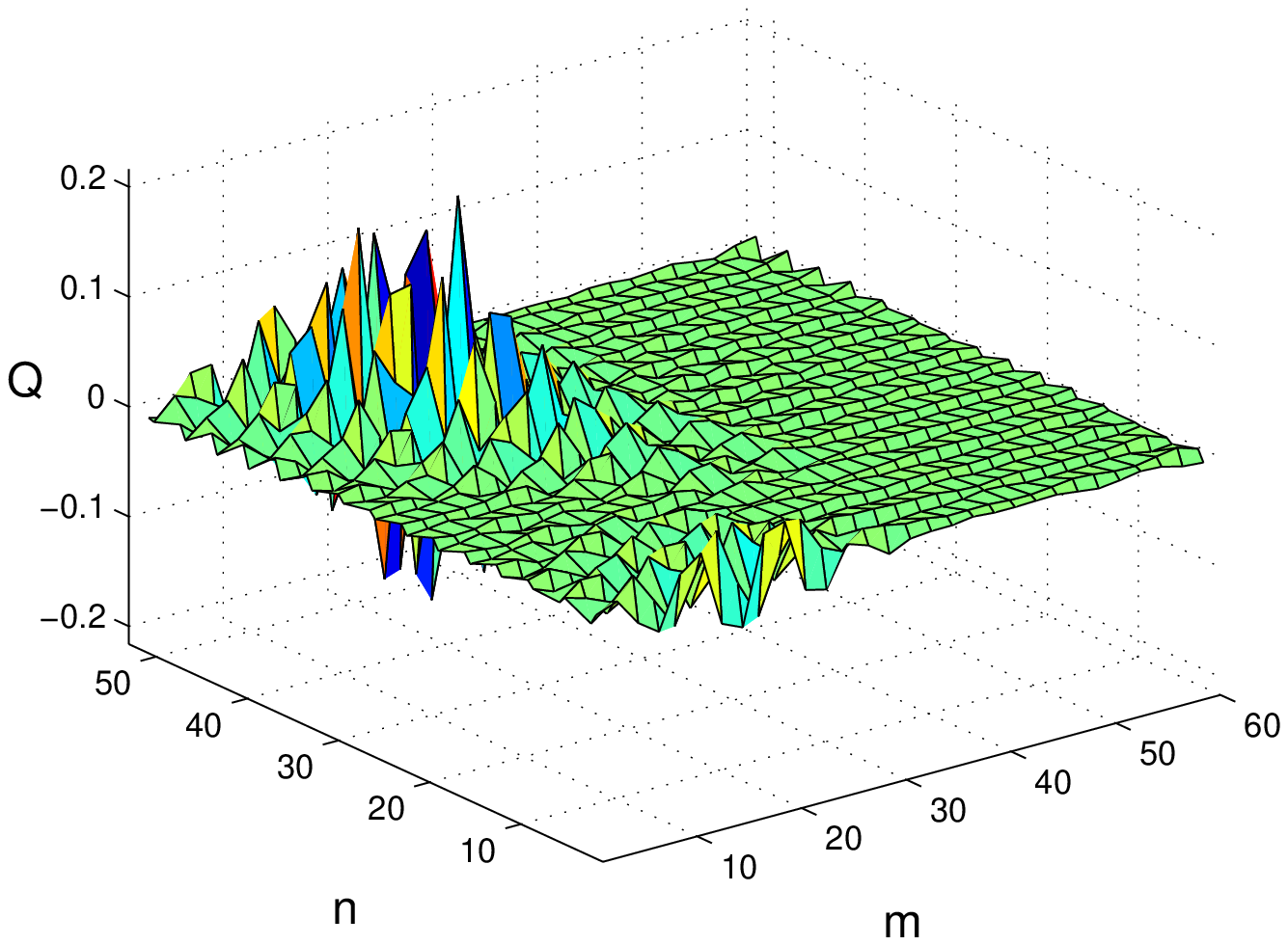}}\\
\vspace{.3in} 
\subfigure[Profile at $t=90$, $\elo = 0.5939$.]{
\includegraphics[width=.45\textwidth]{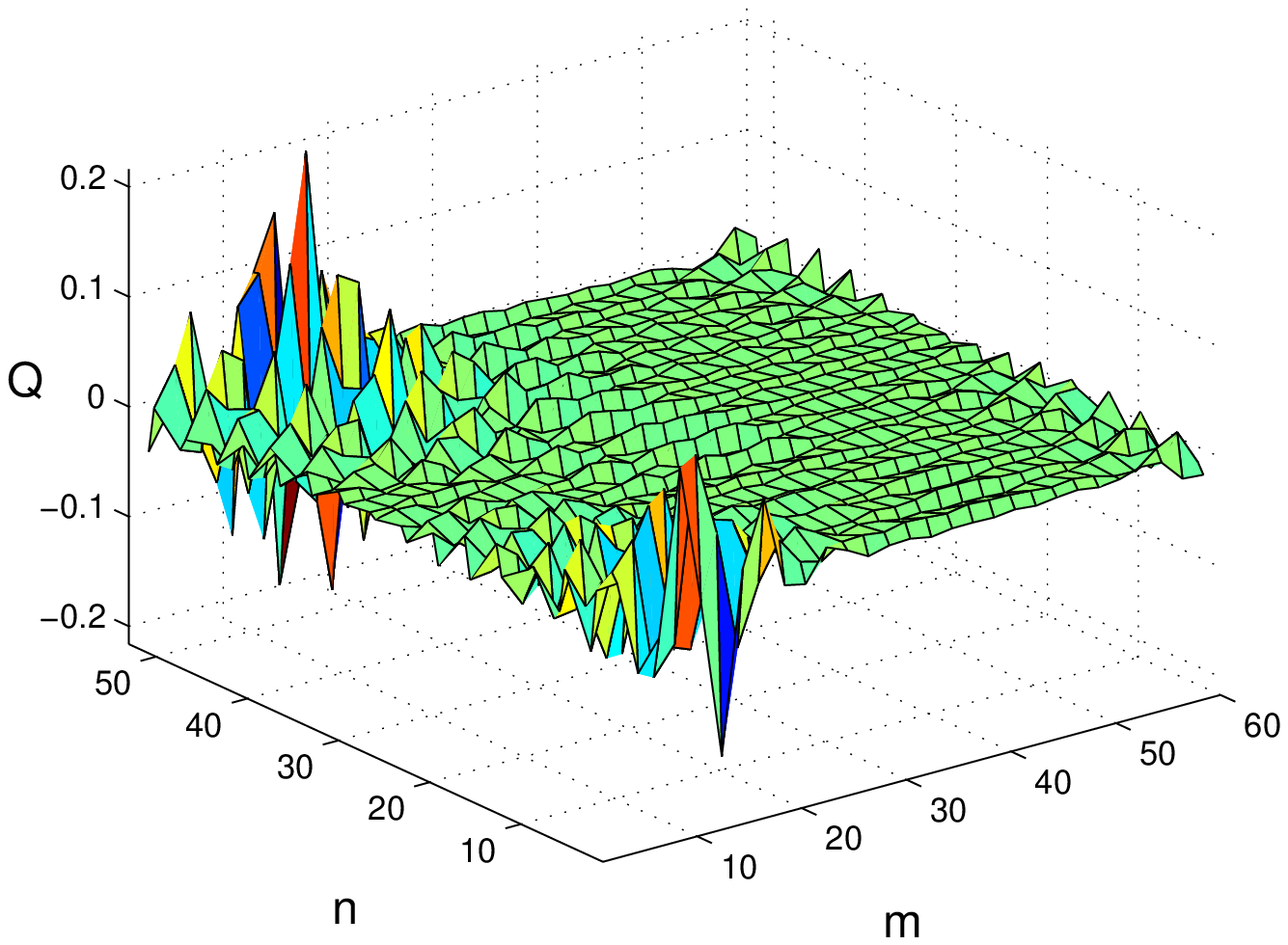}}
\hspace{.3in} 
\subfigure[Profile at $t=120$, $\elo = 0.5897$.]{
\includegraphics[width=.45\textwidth]{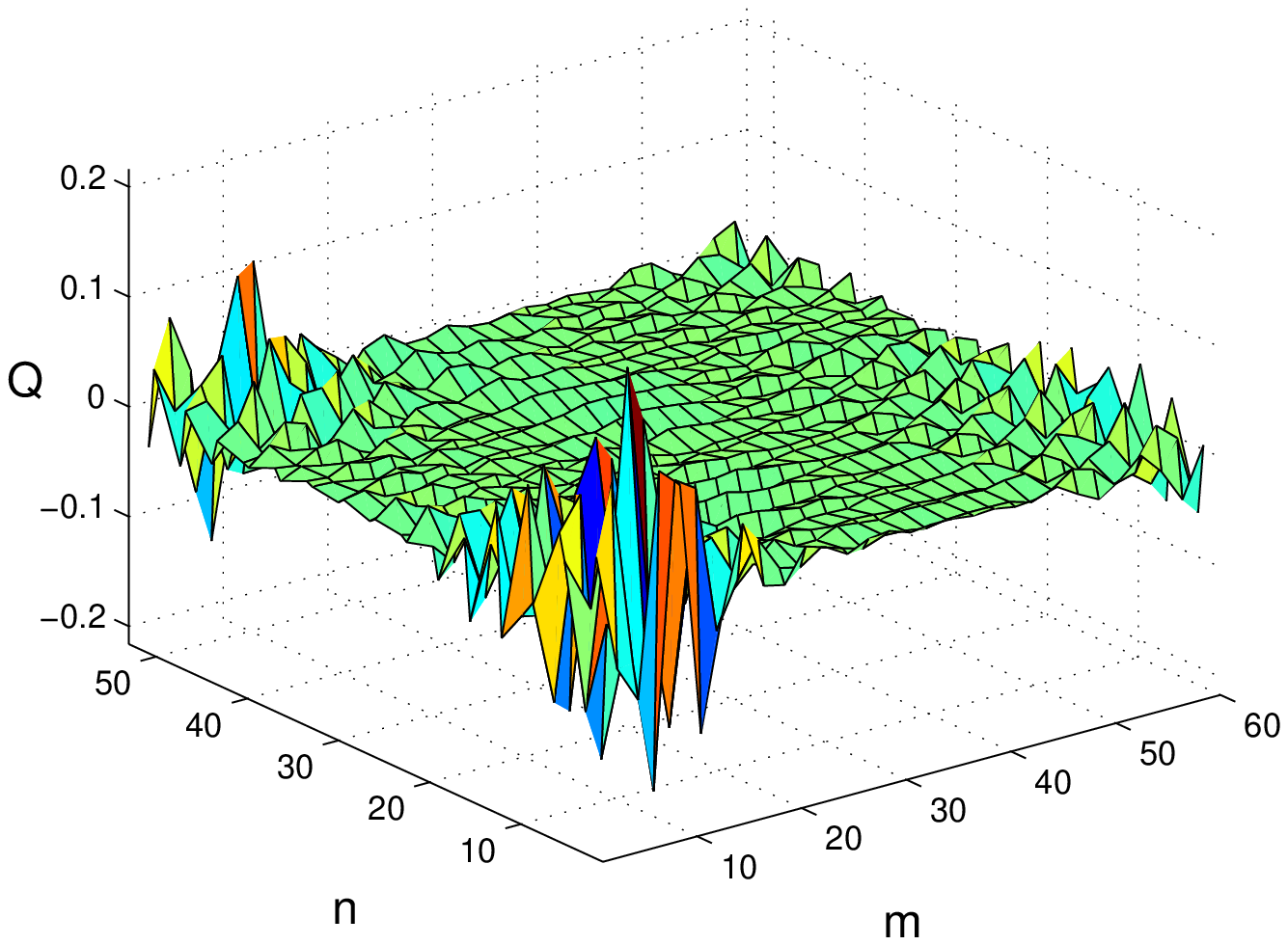}}
\caption{Breather moving at $\Psi=130\degrees$, see Section 
\ref{hpsi130} for details.}
\label{simh7}
\end{figure}

We have presented simulations of breathers which move along axes of 
symmetry of the lattice.  In \Sref{hexdell}, from the results of our asymptotic 
analysis, we found that breather solutions could be constructed for any 
direction of travel.  In this section, we test the mobility of breathers along 
directions which do not correspond to axes of symmetry of the lattice.
We have successfully propagated breathers in a range of directions, 
and here we show only one such breather, moving at an angle
$\Psi=130\degrees$.  We set $k=0.8$ and $l=1.9987$, from which we 
find  $u=-0.2160$, $v=0.2575$, $\Psi = 130\degrees$, $\w=2.9502$ 
and the period is $T=2.1298$.  The remaining parameters are $b=1$, 
$d=1$, $N=30$, $\eee=0.1$, $\lambda=1$, $\alpha=1.0297$ and 
$\beta=1.8514$.  From Figure \ref{simh7}, 
Even at 120 seconds, the breather remains localised without suffering
appreciable degradation nor is much radiation left behind in its wake.   
Initially the energy $E_0$ is computed as 0.5977, only 0.2\% different 
from the asymptotic estimate of $E_0=0.5965$.   
After 120 seconds, $E_0$ is 0.5897, a change of 1.3\%.  
To find the velocity of the breather, we have recorded its motion
and presented relevant data in \Tref{tsimh7}.  From (\ref{velsuv}), 
the average speed is predicted to be 0.3254 units per second; 
whilst our numerical simulation shows an average speed of 0.3361 
units per second,  3.2\% lower than the expected speed.  The 
direction of motion of the breather is predicted accurately.

\begin{table}
\begin{center}
\begin{tabular}[htbp]{| c | c | c | c | c | c | c |} \hline 
{\ssz{Time}} & {\ssz{Horizontal}} & {\ssz{Vertical}} & 
{\ssz{Average horizontal}} & {\ssz{Average vertical}} & {} & {} \\[-8pt] 
{\ssz{(s)}} & {\ssz{displacement}} & {\ssz{displacement}} & 
{\ssz{velocity (units s$^{-1}$)}} & {\ssz{velocity (units s$^{-1}$)}} & 
{\ssz{$\tan\Psi$}} & {\ssz{$\Psi$}} \\ \hline 
{\ssz{$30$}} & {\ssz{$-6$}}  & {\ssz{$7$}}  & {\ssz{$-0.2$}}  & {\ssz{$0.2333$}}  & {\ssz{$-1.1667$}} & {\ssz{$130.60\degrees$}} \\ \hline 
{\ssz{$60$}} & {\ssz{$-11$}}  & {\ssz{$13.5$}}  & {\ssz{$-0.1833$}}  & {\ssz{$0.225$}}  & {\ssz{$-1.2273$}} & {\ssz{$129.17\degrees$}} \\ \hline 
{\ssz{$90$}} & {\ssz{$-17.5$}}  & {\ssz{$21.5$}}  & {\ssz{$-0.1944$}}  & {\ssz{$0.2389$}}  & {\ssz{$-1.2286$}} & {\ssz{$129.14\degrees$}} \\ \hline  
{\ssz{$120$}} & {\ssz{$-25$}}  & {\ssz{$30$}}  & {\ssz{$-0.2083$}}  & {\ssz{$0.25$}}  & {\ssz{$-1.2$}} & {\ssz{$129.81\degrees$}} \\ \hline
\end{tabular}
\end{center}
\caption{Summary of breather motion ($\Psi=130\degrees$) 
see Section \ref{hpsi130} for details.}
\label{tsimh7}
\end{table}

\subsection{Stationary breather in a lattice with asymmetric potential} 
\lbl{hnra}

In Sections \ref{hnstsy}--\ref{hpsi130}, we have shown simulations
of lattices with a symmetric potential, that is, with $a$ and $c$ not 
necessarily zero in \eref{heqqfo}.   We now consider the more general 
case for which the interaction potential is asymmetric ($a\neq0 \neq c$).  
To illustrate a stationary breather, we generate initial data using 
\eref{hexqasym},  with the wavevector $\mb{k} = \mb{k_1} = 
[\pi/3,\pi/h]^T$.  As discussed in \Sref{hexasymm}, anomalous dispersion 
corresponds to $b>10a^2/9$, hence we choose $a=1$, $b=2.5$, $c=0$ 
and $d=1$.  The remaining parameter values are $N=30$, $\eee=0.1$, 
$\lambda=1$, hence $\w=3$, $T=2.0668$, $\alpha=0.8665$ and $\beta=1.8670$.

\begin{figure}[htbp]
\centering 
\subfigure[$t=10T=20.69$.]{
\includegraphics[width=.45\textwidth]{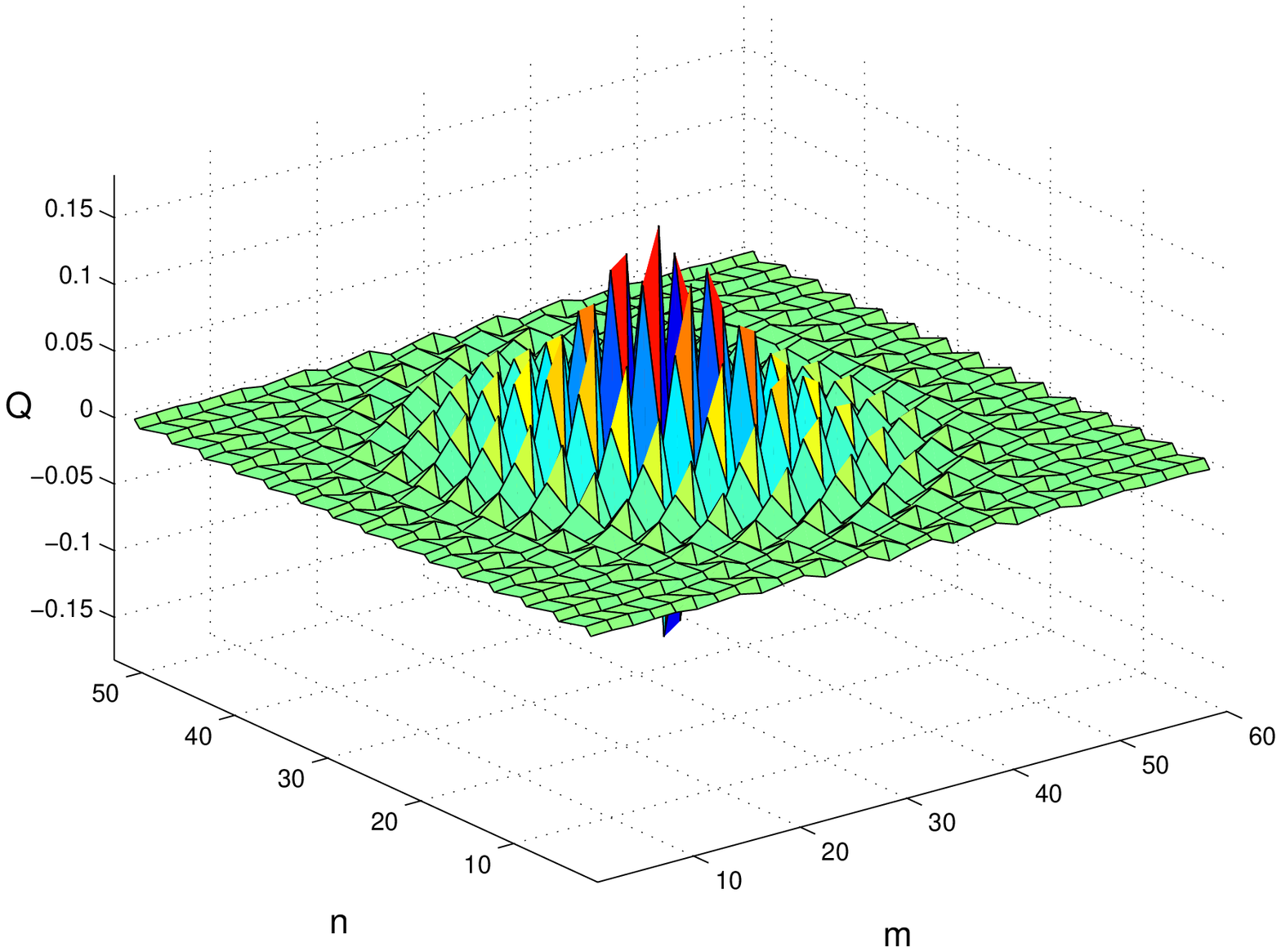}}
\hspace{.3in} 
\subfigure[Plot of $e_{m,n}$, $E_0=0.5453$.]{
\includegraphics[width=.45\textwidth]{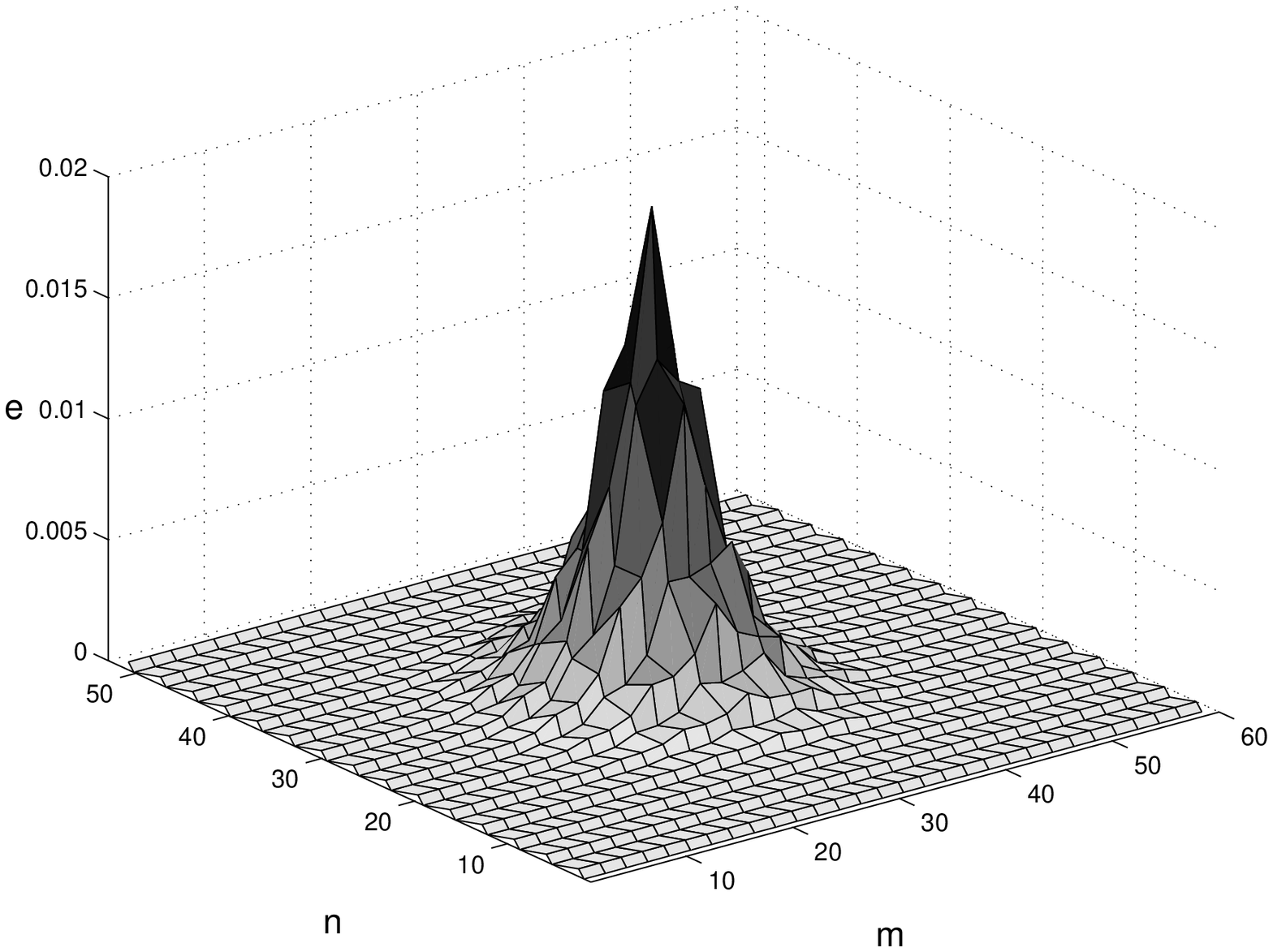}} \\
\vspace{.3in} 
\subfigure[$t=30T=41.34$.]{
\includegraphics[width=.45\textwidth]{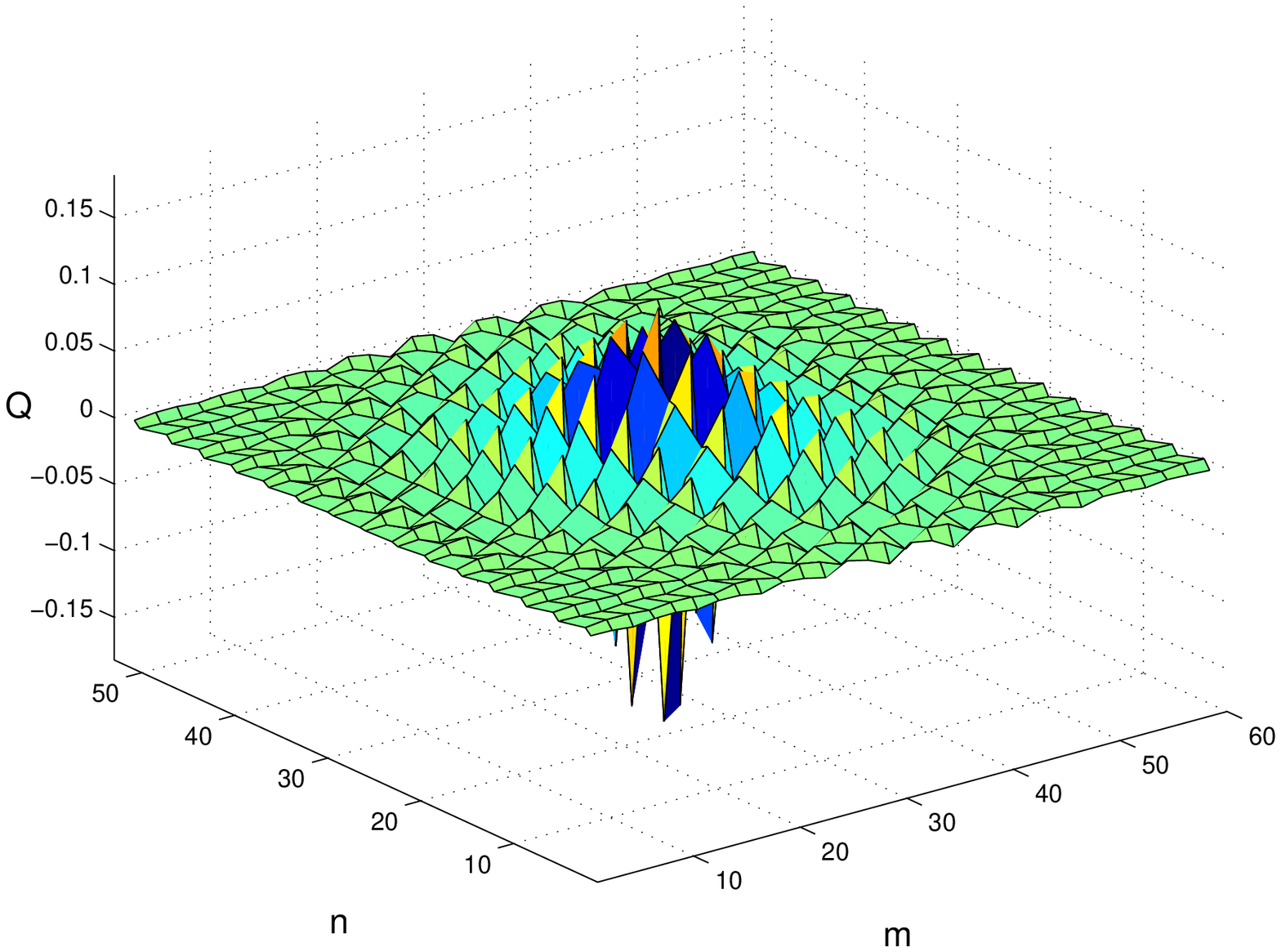}}
\hspace{.3in} 
\subfigure[Plot of $e_{m,n}$, $E_0=0.5425$.]{
\includegraphics[width=.45\textwidth]{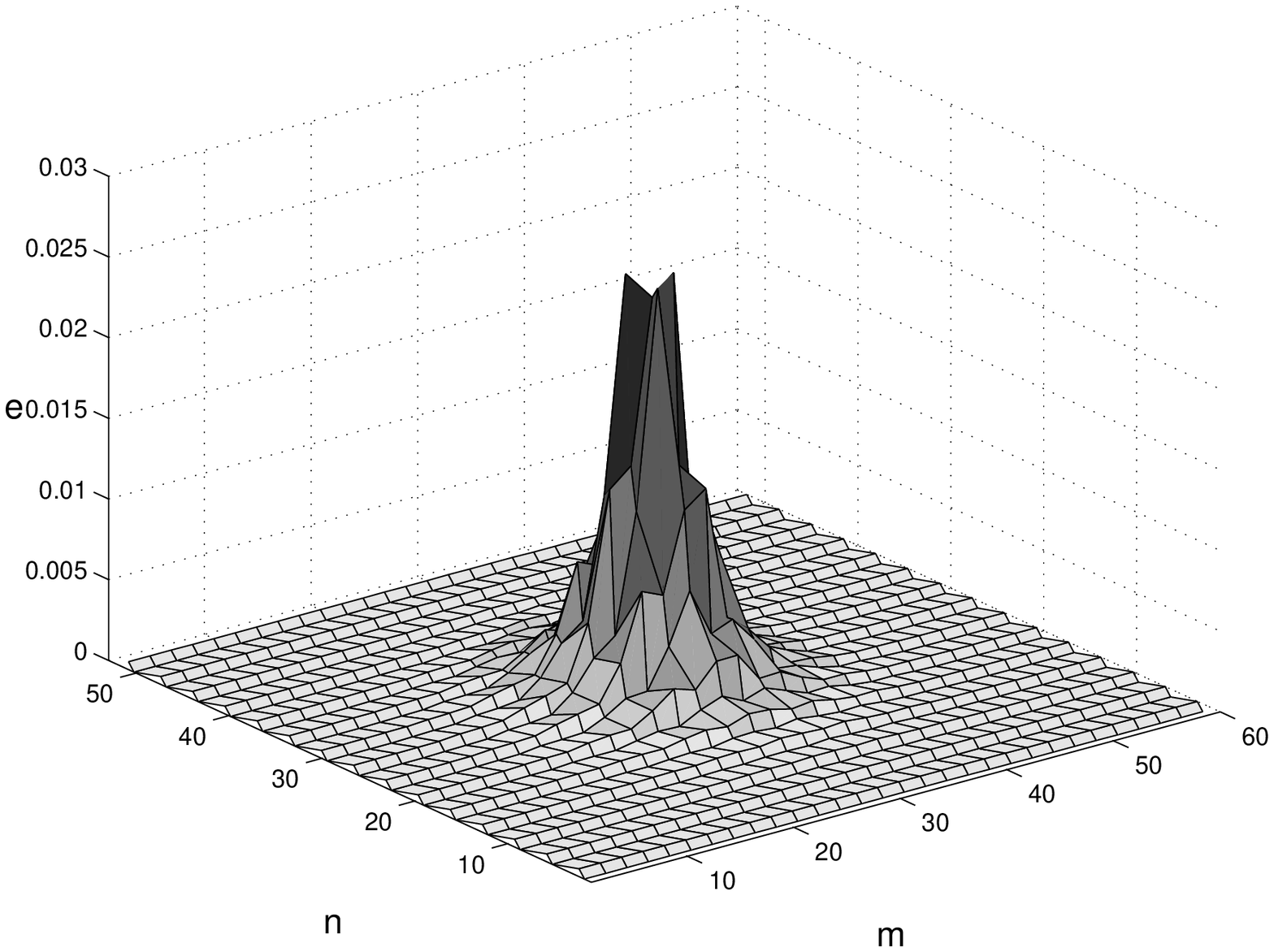}} \\
\vspace{.3in} 
\subfigure[$t=40T=82.67$.]{
\includegraphics[width=.45\textwidth]{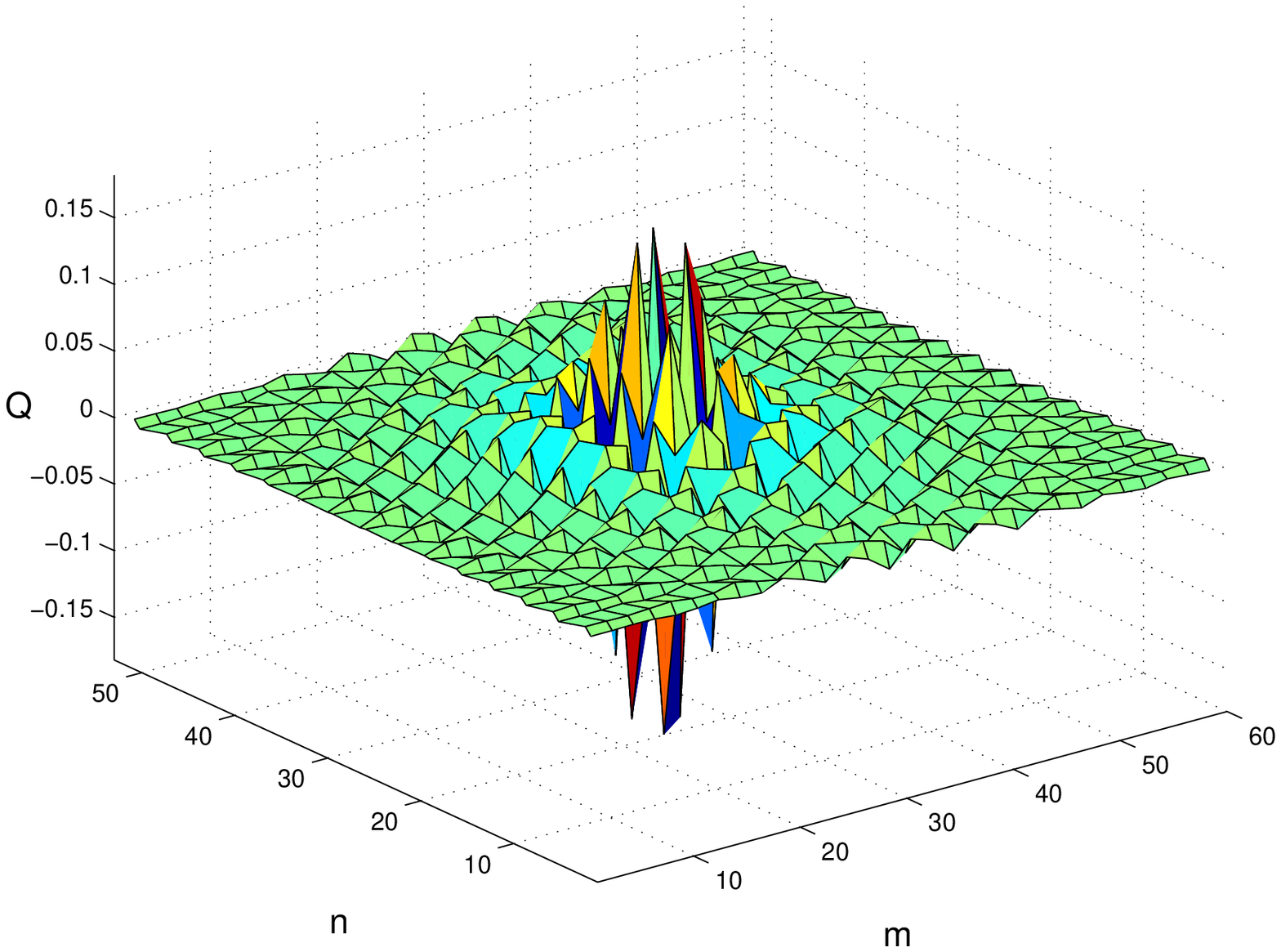}}
\hspace{.3in} 
\subfigure[Plot of $e_{m,n}$, $E_0=0.5371$.]{
\includegraphics[width=.45\textwidth]{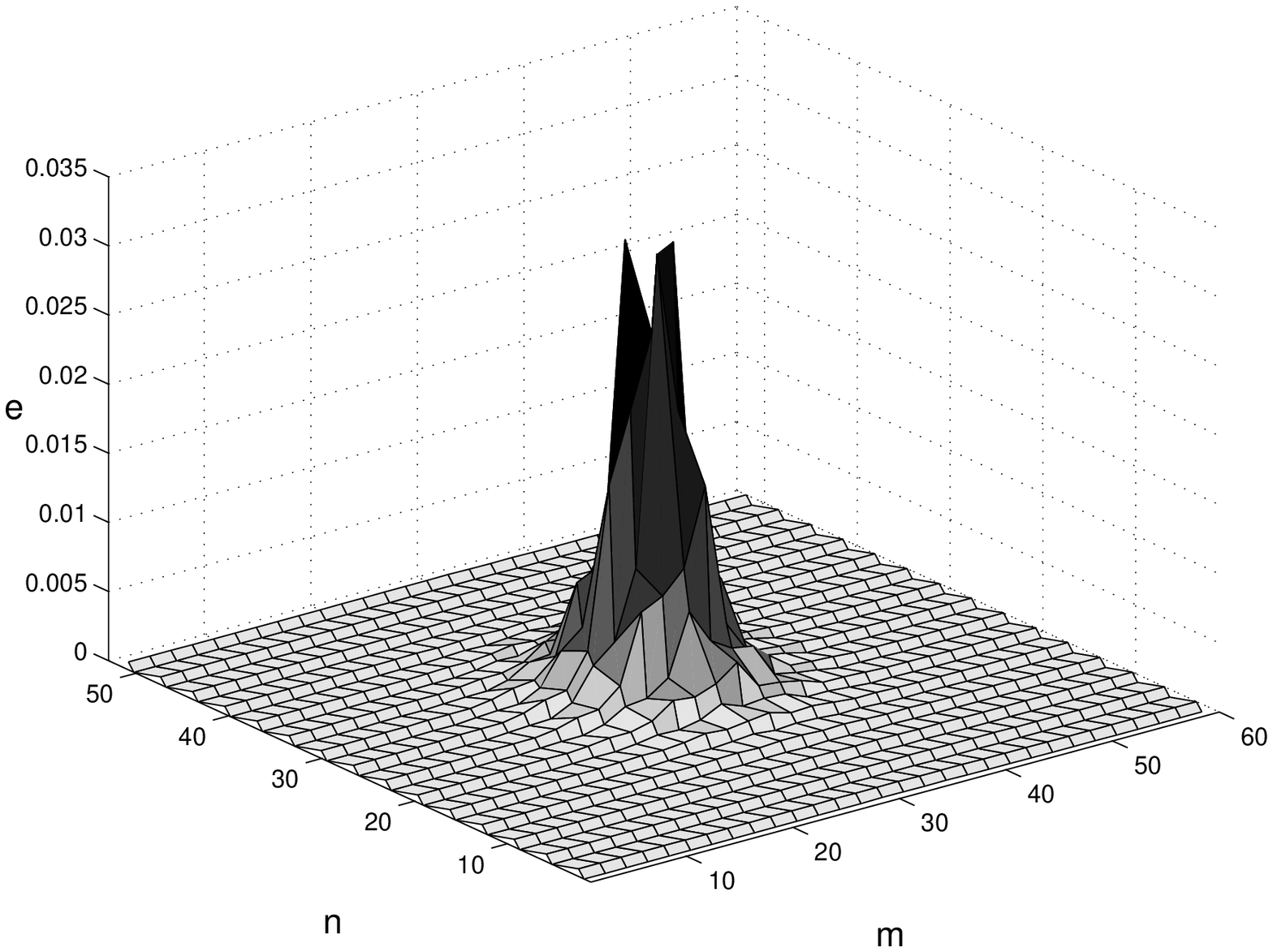}}
\caption{Stationary breather in a lattice with an asymmetric potential, 
see Section \ref{hnra} for details.}
\label{simh5}
\end{figure}

The breather is shown in \Fref{simh5} after $10$, $30$ and $40$ full 
oscillations.  Initially, we find the breather's width and energy are  
$\wbr=7.46$ and $\elo=0.5429$, the latter being only 0.26\% different 
from the asymptotic estimate \eref{henex} of 0.5416.   The accompanying 
plots of $e_{m,n}$ demonstrate clearly that the breather preserves its 
form and remains localised, even after 80 seconds.  At $t=40T$, we find 
that $\wbr=6.83$, a narrowing of the breather by 8\%.  Also, at $t=40T$, 
the numerically computed value of the energy $E_0=0.5371$, showing 
that the energy does not fluctuate significantly.

\section{Discussion} \lbl{hdisc}

In this paper, we have found approximations to discrete breathers
in a two-dimensional hexagonal FPU lattice with a scalar-valued 
function at each node.  We have shown that
the lattice equations can be reduced to a cubic NLS equation for
two special cases.  In \Sref{hexsym}, we obtained moving
breathers when the interaction potential is symmetric; in
this case an ellipticity criterion for the wavevector is found.
In the appendix we summarise our technique for approximating 
soliton solutions of the two-dimensional NLS equation.

In \Sref{hexasymm}, we considered lattices with an asymmetric 
potential, in which case a reduction to NLS could only be performed
for stationary breathers.   
The theoretical methods employed here are similar to those used on 
the square lattice of \citep{buts06}; however,  several important 
differences emerge in the course of the analysis.  
We found that the anomalous dispersion regime corresponds to 
$b>4a^2/3$ in the square lattice and $b>10a^2/9$ in the hexagonal.  
Furthermore we find that stationary breathers involve the 
generation of second harmonics in the hexagonal lattice 
($G_2=aF^2/3$), whereas they are suppressed in the square 
lattice ($G_2=0$).  In both lattices, the quadratic nonlinearity 
generates a small amplitude slowly-varying mode ($G_0=-a|F|^2$).

For symmetric interactions we found an associated ellipticity constraint: 
moving breathers only occur for certain wavevectors, this means that 
all breathers have (i) a relatively high frequency, and (ii) a maximum 
speed, which depends on the wave vector and hence on the 
direction of travel, (iii) a threshold energy which also depends on 
wavevector and hence is related to speed and direction of travel.  
We find that the threshold energy is lower for breathers at the 
edge of the domain of ellipticity in wavevector space. These breathers 
have lower frequencies and faster speeds.  The fastest-moving 
breathers are restricted to moving along lattice directions.

We also presented asymptotic estimates for the breather energy in
Sections \ref{hexsym} and \ref{hexasymm}.  As expected, we found a
minimum energy required to create breathers in the hexagonal
lattice.  The threshold energy for moving breathers is smaller
than that required for stationary breathers and becomes vanishingly
small at the boundary of the domain of ellipticity.

In \Sref{hex5}, we extended the small amplitude expansion to fifth-order 
and derived a higher-order equation which more correctly describes 
the shape and stability properties of the breather envelope.  We 
obtained a generalised NLS equation \eref{hexgennls} with a variety 
of perturbation terms, some of which are known to be stabilising.  This 
equation is slightly simpler than the corresponding equation obtained 
for the square lattice, namely (3.11) of \citep{buts06}.  In particular, 
the higher-order dispersive terms in \eref{hexgennls} are isotropic, 
reflecting the hexagonal rotational symmetry of the lattice.  For stationary 
breathers in the case of a symmetric potential we find that the cubic 
nonlinearity does not give rise to third harmonics, that is, $H_3=0$ in 
the hexagonal lattice, in contrast to $H_3=bF^3/8$ in the square lattice.

In \Sref{hexnumerics}, we illustrated  these breather modes, showing 
that both stationary and moving breathers are long-lived.  The 
breather profiles change little over time, a small amount of 
energy is shed due to the initial conditions being only approximate.  
We have successfully propagated long-lived breathers moving in 
directions which are not axes of symmetry of the lattice (for 
instance, $\Psi=130\degrees$),  suggesting that there is no absolute 
restriction upon the direction of travel; this is in contrast to the 
observations reported by Marin \etal \citep{mar00, mar98} for 
mechanical (two-component) two-dimensional lattices, who could 
only find breathers which travelled along axes of symmetry of the lattice.

\ack

IAB would like to thank both the UK Engineering and Physical Sciences 
Research Council for financial assistance, and also Qamran Yaqoob for his 
assistance in the preparation of several diagrams that appear in this paper.

\appendix 
\renewcommand{\theequation}{\Alph{section}\arabic{equation}}
\section{Approximation to Townes soliton} \lbl{app}

Since analytic formulae for Townes solitons are unavailable, we use the 
Rayleigh-Ritz method to find time-harmonic radially symmetric solutions of
\begin{equation}
\ii F_T + D \nabla^2 F + B |F|^2 F= 0 ,  
\lbl{app-pde} 
\end{equation}
of the form $F({\bf x},T)=e^{\ii\lambda T}\phi(r)$ where 
$r=|{\bf x}|=\sqrt{\xi^2 + \eta^2}$.   The function $\phi$ satisfies 
\begin{equation} \lbl{nlsrl}
-\lambda \phi + D\nabla^2 \phi + B\phi^3=0 ,
\end{equation}
where $\nabla^2=\partial_\xi^2 + \partial_\eta^2$.
Equation (\ref{nlsrl}) arises from a variational derivative of  
\begin{equation} \lbl{hvar}
\mcal{E}(\phi) = \int \int \shalf \lambda |\phi|^2 +
\shalf D |\nabla \phi|^2 - \sfr{1}{4} B |\phi|^4 d^2{\bf r}.
\end{equation}
Using a trial solution of the form $\phi=\alpha\,\sech(\beta r)$, 
where $\alpha$ and $\beta$ are undetermined parameters, we find 
\begin{equation} \lbl{hab}
\mcal{E}(\alpha,\beta) = \frac{D(1+2\str{ln}2)}{12} \alpha^2 -
\frac{B(4\str{ln}2-1)}{24}\frac{\alpha^4}{\beta^2} + \frac{\lambda
\str{ln}2}{2}\frac{\alpha^2}{\beta^2} ; 
\end{equation}
$\alpha$ and $\beta$ are determined by seeking stationary 
points of the action $\mcal{E}$, namely  ${\partial \mcal{E}}/
{ \partial \alpha} = {\partial \mcal{E}}/{ \partial \beta} = 0$.
Hence we find 
\begin{equation} \lbl{alpbet}
\alpha = \sqrt{\frac{12\lambda\str{ln}2}{B(4\str{ln}2 -1)}} , \qquad 
\beta = \sqrt{\frac{6\lambda\str{ln}2}{D(2\str{ln}2 + 1)}} , 
\end{equation}
and so our approximation to the Townes soliton solution of (\ref{app-pde}) is
\begin{equation} \lbl{vartown}
F = \sqrt{\frac{12\lambda\log2}{B(4\log2 -1)}} \ \exp (\ii \lambda T) \
\sech \left( \sqrt{\frac{6\lambda\log 2}{D(2\log2 + 1)}}
\sqrt{\xi^2 + \eta^2}\right) .
\end{equation}

\footnotesize

\end{document}